\documentclass[article]{aa}

\usepackage{graphicx}
\usepackage{amsmath}

\usepackage{txfonts}
\usepackage{hyperref}
\usepackage{natbib}
\bibpunct{(}{)}{;}{a}{}{,} 
\usepackage{color}
\usepackage{xspace}

\usepackage{xcolor}

\usepackage{listings}

\definecolor{orange}{rgb}{1.0, 0.4980392156862745, 0.054901960784313725}
\definecolor{codegreen}{rgb}{0,0.6,0}
\definecolor{codegray}{rgb}{0.5,0.5,0.5}
\definecolor{codepurple}{rgb}{0.58,0,0.82}
\definecolor{backcolour}{rgb}{0.95,0.95,0.92}

\lstdefinestyle{mystyle}{
    backgroundcolor=\color{backcolour},
    commentstyle=\color{codegreen},
    keywordstyle=\color{magenta},
    numberstyle=\tiny\color{codegray},
    stringstyle=\color{codepurple},
    basicstyle=\ttfamily\footnotesize,
    breakatwhitespace=false,
    breaklines=true,
    captionpos=b,
    keepspaces=true,
    numbers=left,
    numbersep=5pt,
    showspaces=false,
    showstringspaces=false,
    showtabs=false,
    tabsize=2
}

\newcommand{\Gaia}{{\it Gaia}\xspace}
\newcommand{\cat}{\ensuremath{\mathcal{C}}}
\newcommand{\vq}{\ensuremath{\mathbf{q}}}

\newcommand{\cca}{Center for Computational Astrophysics, Flatiron Institute, 162 Fifth Ave, New York, NY 10010, USA}
\newcommand{\ccpp}{Center for Cosmology and Particle Physics, Department of Physics, New~York~University, 726~Broadway, New~York, NY 10003, USA}
\newcommand{\mpia}{Max-Planck-Institut f\"ur Astronomie, K\"onigstuhl 17, D-69117 Heidelberg, Germany}
\newcommand{\leiden}{Leiden Observatory, Leiden University, Niels Bohrweg 2, 2333 CA Leiden, The Netherlands}
\newcommand{\cambridge}{Institute of Astronomy, University of Cambridge, Madingley Road, Cambridge CB3 0HA, United Kingdom}

\newcommand{\torino}{INAF - Osservatorio Astrofisico di Torino, Strada Osservatorio 20, Pino Torinese 10025 Torino, Italy}

\newcommand{\harvard}{Department of Physics, Harvard University, 17 Oxford St., Cambridge, MA 02138, USA}
\newcommand{\CFA}{Harvard-Smithsonian Center for Astrophysics, 60 Garden St., Cambridge, MA 02138, USA}
\newcommand{\Monash}{School of Physics \& Astronomy, Monash University, Clayton 3800, Victoria, Australia}
\newcommand{\AstroTD}{Centre of Excellence for Astrophysics in Three Dimensions (ASTRO-3D), Melbourne, Victoria, Australia}
\begin{document}

   \title{An empirical model of the \Gaia DR3 selection function}

    \titlerunning{\Gaia DR3 selection function}
    \authorrunning{Cantat-Gaudin et al.}

   \author{Tristan Cantat-Gaudin\inst{1} 
          \and
          Morgan Fouesneau\inst{1}  
          \and
          Hans-Walter Rix\inst{1} 
          \and
          Anthony G. A. Brown\inst{2} 
          \and
            Alfred Castro-Ginard\inst{2} 
            \and
            Ronald Drimmel\inst{3} 
            \and
            David W. Hogg\inst{1,4,5} 
            \and
            Andrew R. Casey\inst{7,8} 
            \and
            Shourya Khanna\inst{3} 
            \and
            Semyeong Oh\inst{6} 
            \and
            Adrian~M.~Price-Whelan\inst{4} 
            \and
            Vasily Belokurov\inst{6} 
            \and
            Andrew K. Saydjari\inst{9,10} 
            \and
            Gregory M. Green\inst{1}
          }

   \institute{
        \mpia  \\ \email{cantat@mpia.de}
         \and \leiden \and \torino \and \cca \and \ccpp \and \cambridge \and \Monash \and \AstroTD \and \harvard \and \CFA 
             }

   \date{}



  \abstract{Interpreting and modelling astronomical catalogues requires an understanding of the catalogues' completeness or selection function: objects of what properties had a chance to end up in the catalogue. Here we set out to empirically quantify the completeness of the overall \Gaia DR3 catalogue. This task is not straightforward because \Gaia is the all-sky optical survey with the highest angular resolution to date and no consistent ``ground truth'' exists to allow direct comparisons.
  However, well-characterised deeper imaging enables an empirical assessment of \Gaia's $G$-band completeness across parts of the sky.
  On this basis, we devised a simple analytical completeness model of \Gaia as a function of the observed $G$ magnitude and position over the sky, which accounts for both the effects of crowding and the complex \Gaia scanning law. Our model only depends on a single quantity: the median magnitude $M_{10}$ in a patch of the sky of catalogued sources with \texttt{astrometric\_matched\_transits} $\leq 10$.  $M_{10}$ reflects elementary completeness decisions in the \Gaia pipeline and is computable from the \Gaia DR3 catalogue itself and therefore applicable across the whole sky. We calibrate our model using the Dark Energy Camera Plane Survey (DECaPS) and test its predictions against Hubble Space Telescope observations of globular clusters. We find that our model predicts \Gaia's completeness values to a few per cent ({\it RMS}) across the sky. We make the model available as a part of the \texttt{gaiasf} Python package built and maintained by the GaiaUnlimited project\thanks{\url{https://github.com/gaia-unlimited/gaiasf}}. }


   \maketitle
%
\section{Introduction}\label{sec:intro}
Comparing model predictions to observations drawn from an astronomical catalogue requires knowledge of the selection effects and incompleteness affecting the observed list of objects. Knowing what we could not observe could be as essential as what we observed, even for simple endeavours. For instance, mapping the stellar distribution around the Sun to reconstruct our Galaxy's overall shape requires knowing the sample's limiting magnitude and whether this limit varies across the sky.

A catalogue {\it selection function} $S_\cat$ describes the probability of an object to be included in an astronomical catalogue. Such a function represents the combined effects of the data collection (such as detection efficiency decreasing with apparent magnitude) and data processing (such as removing sources with noisy observations).
To avoid biases caused by incomplete data, astronomers commonly restrict their studies to regions of the parameter space where the sample is assumed to be complete ($S_\cat \sim 1$). This approach is generally substantially restrictive and could lead to a poor representation of the problem one wants to address. Instead, one needs to fold in the selection function of a catalogue. In \citet{Rix21}, we presented a general approach to evaluating and \textit{accounting} for known selection functions in modelling astronomical data.

\Gaia observes the sky continuously according to a complex scanning law: a six-hour rotation around its spin axis, a 63-day day precession of the spin axis, and the annual motion of the Earth (and its Lagrange 2 point) around the Sun \citep{gaiaMission}. This results in an intricate pattern, covering the entire celestial sphere with on average $\sim15$ visits per year, but with significant variations. 
The \Gaia catalogue, only includes sources with at least five observations \citep{Lindegren18}. The probability that a transit across the \Gaia field of view leads to an observation is lower for fainter sources, mainly due to two reasons.
First, the onboard source detection algorithm has a nominal faint-end threshold of $G=20.7$\,mag beyond which sources are not selected for observation, but the onboard magnitude estimate has a precision of a few tens of magnitude \citep{deBruijne15}, giving sources fainter than 20.7 a non-zero observation probability. Second, in crowded areas where the density exceeds $\sim$1,050,000 per square degree, \Gaia cannot follow all transiting sources and prioritises bright objects over fainter ones \citep{gaiaMission}. The probability that a source benefits from five observations is therefore a complex function of sky position (via both crowding and the scanning law) and magnitude. Using the notation introduced in \citet{Rix21}, this paper models the catalogue selection function $S^{\rm parent}(\vq)$ of \Gaia~DR3, where in the present case the catalogue properties $\vq=(\ell,b,G)$ are the sky-position $(\ell, b)$ and $G$ magnitude of a source.

The most common and straightforward approach to estimating a catalogue selection function of a sample is to compare the dataset with a more complete catalogue, which often means a deeper one in terms of magnitude limit. This comparison is generally made by binning both catalogues by magnitude, colour, or sky position and computing the ratio of source count in each bin.
For instance, \citet{Rybizki21rvs} followed this procedure to characterise the selection function of the \Gaia~DR2 Radial Velocity sample \citep{Katz19}, and \citet{Everall20seestar} improved upon this version using a smooth Gaussian Mixture Model to solve the issue with sparsely-populated bins. However, this technique is empirical and relies on an external reference of complete samples.

There are ongoing efforts to reconstruct the \Gaia selection function from a forward-modelling approach \citep[][and subsequent papers in their \textit{Completeness of the Gaiaverse} series]{Boubert20gaiaverse1, Boubert20gaiaverse2}. This approach requires modelling each step of the \Gaia processing, from the scanning law and onboard filtering to the astrometric processing. In an upcoming paper (Castro-Ginard et al., in prep.), we will update their model, using transit data of non-variable stars to identify data-taking gaps and time-variable detection efficiencies and to extend the model to bright sources ($G\sim$ 1 to 6\,mag).

This study is part of the larger GaiaUnlimited project\footnote{\url{https://gaia-unlimited.org/}}, which aims to determine the \Gaia selection function and provide tools to the astronomical community to account for the selection effects in the \Gaia catalogue. The present paper empirically builds an analytical model of the \Gaia DR3 source catalogue selection function, i.e., the probability that the final catalogue contains a given source as a function of its sky position and an apparent $G$ magnitude. We use the deep Dark Energy Camera Plane Survey \citep[DECaPS][]{Schlafly18,Saydjari22} of the southern Galactic plane as our ``complete'' reference to calibrate our model. We identified a simple quantity derived from the \Gaia catalogue itself to use as a predictor of the \Gaia completeness as a function of magnitude at any location over the sky, even outside the DECaPS footprint.

The approach and modelling are presented in Sect.~\ref{sec:approach}. In Sect.~\ref{sec:testing} we verify our predictions against data which were not used in the model calibration. We discuss our model and its limitation in Sect.~\ref{sec:discussion}, and close with concluding remarks in Sect.~\ref{sec:conclusion}.

\section{Approach} \label{sec:approach}

\subsection{Choice of reference datasets} \label{sec:data}

We use the Dark Energy Camera Plane Survey (DECaPS) DR1 catalogue \citep[][]{Schlafly18} as ground truth to calibrate our model of the \Gaia selection function. DECaPS is a ground-based optical and near-infrared survey of the Galactic plane using the Dark Energy Camera (DECam, \citealt{Flaugher15}) mounted on the 4m Victor M. Blanco telescope at the Cerro Tololo Inter-American Observatory (CTIO). The 2.2$^{\circ}$\ diameter field of view, 0.26$"$/pixel plate scale and arcsecond seeing make these observations well-suited to resolving even the extremely crowded inner galaxy. DECaPS DR1 covers the Galactic plane with $|b|< \sim 4^{\circ}$ and $5 > \ell > -120 ^{\circ}$. The survey reaches typical exposure depths of $\sim23$\,mag in $g$ and $r$ bands, and uses the \texttt{crowdsource} photometric pipeline, which is specifically designed to deal with crowded fields.

In Sect.~\ref{sec:HST} we verify the prediction of our model on high-density regions, using Hubble Space Telescope observations of the inner $3.5 \times 3.5$\,arcmin 26 globular clusters collected by \citet{Sarajedini07}. The data were acquired with the Wide Field Channel of the Advanced Camera for Surveys, with photometry in the F606W and F814W filters, and are essentially complete down to magnitude 25. This data set was used in \citet{Arenou18validation} to visualise the completeness of \Gaia~DR2 and the influence of crowding, but no quantitative model of completeness was proposed.

\subsection{Choice of initial dependencies} \label{sec:dependencies}

\begin{figure*}
    \centering
    \includegraphics[width=0.99\textwidth]{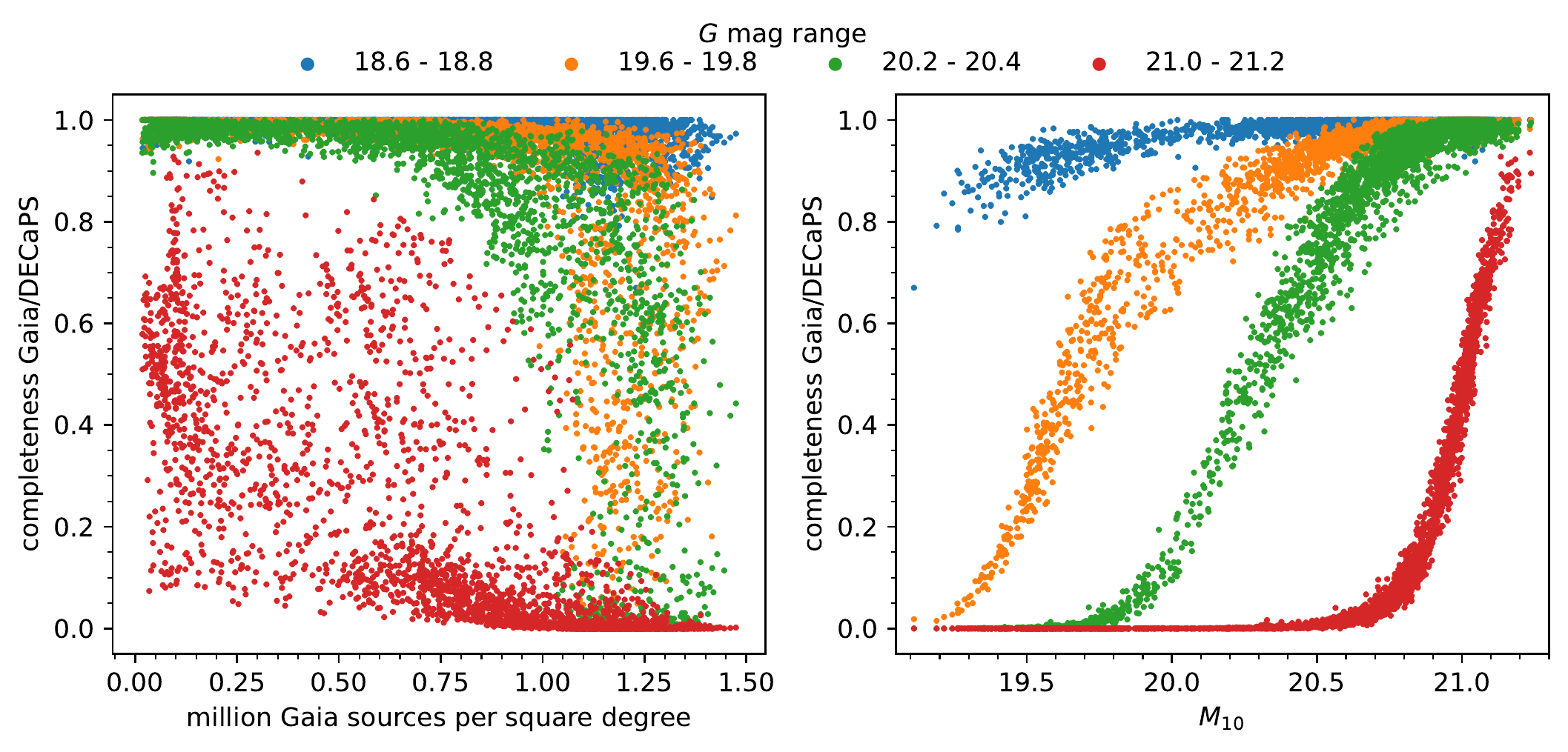}
    \caption{ Completeness of \Gaia relative to the DECaPS survey, which is taken as ``ground truth'', in four magnitude ranges, computed in 3000 distinct patches across the DECaPS footprint. This completeness is shown as a function of \Gaia source density (left) or $M_{10}$ (right):  \Gaia source density is a poor predictor of completeness, while the $M_{10}$ parameter --  the median magnitude of catalogued sources with  \texttt{astrometric\_matched\_transits}$\leq 10$ in a surrounding patch of the sky --  is an excellent completeness predictor. $M_{10}$ combines the impact of source density and scanning law, as demonstrated e.g. in Figure~\ref{fig:mapM10}.
    }
    \label{fig:c_vs_M10_magRanges}
\end{figure*}

This work aims to identify observable quantities that can be computed from the \Gaia data itself, to be used as a proxy to constrain the selection function in any given field. We explored possible choices of observables by computing source count ratios between \Gaia and DECaPS in magnitude bins in various areas on the sky.
Due to the onboard resource allocation strategy prioritising bright sources, a naive expectation would be that completeness correlates with observed source density, with more populated fields being less complete. The observed source density is, in fact, a poor indicator of the true density (two areas with the same number of \Gaia sources can differ by a factor of four in DECaPS) and thus a poor predictor of completeness. This is illustrated in the left panel of Fig.~\ref{fig:c_vs_M10_magRanges}, and in Fig.~\ref{fig:gaia_density_saturation}.

We also tested the following as possible indicators of completeness in a given field of view: the magnitude at which the observed luminosity function differs from an expected power law, the mode of the magnitude distribution, the magnitude of the faintest star in a given area, and the 90th percentile of magnitude. This last quantity provides a reasonable estimate of completeness in the most crowded regions but does not perform well in sparser fields. Constructing an all-sky map of the aforementioned quantities (except local source density) is also computationally very expensive, as it requires going through the entire data set of $\sim$1.8\,billion DR3 sources.

\begin{figure*}
    \centering
    \includegraphics[width=0.99\textwidth]{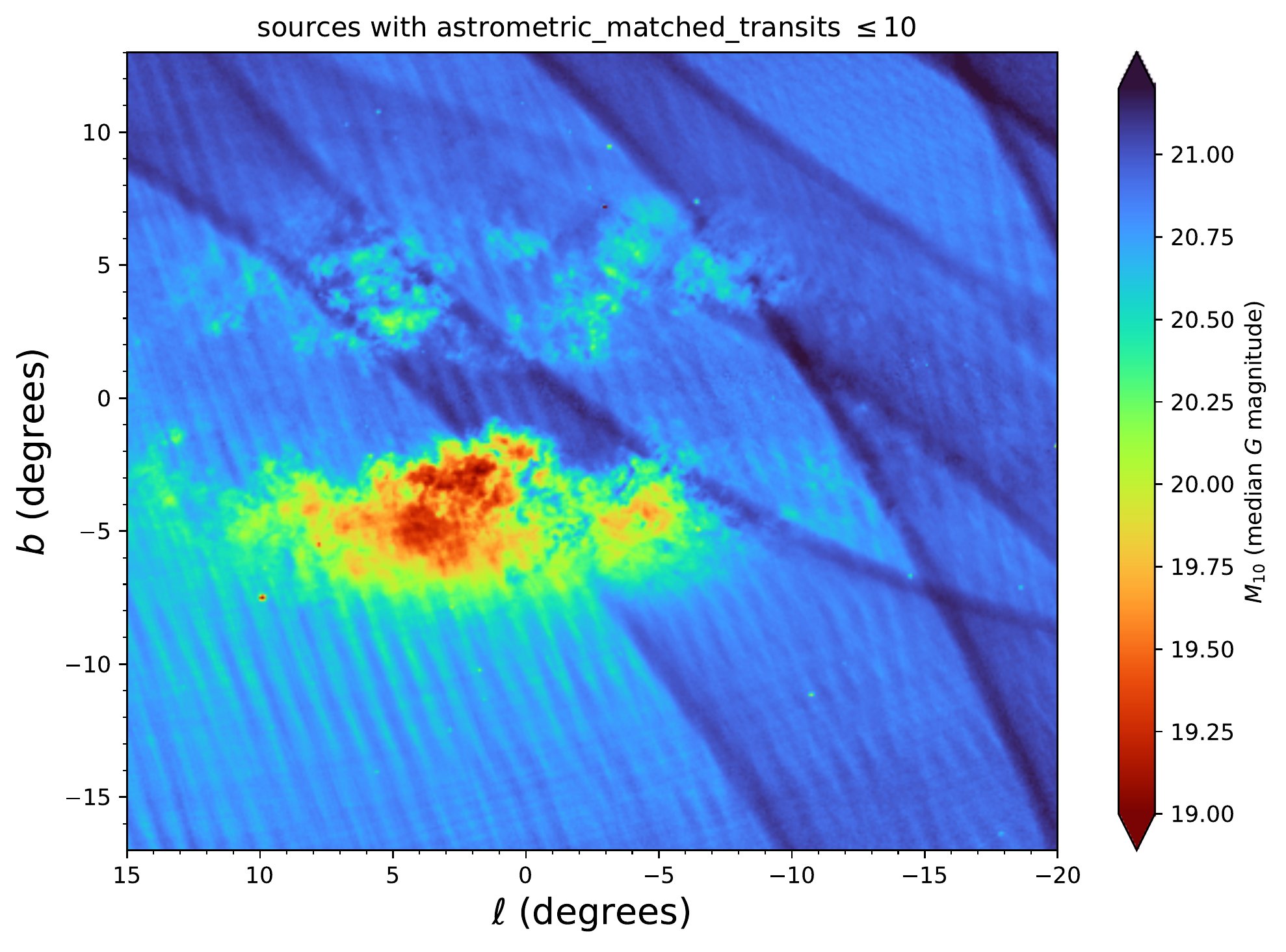}
    \caption{Map of the parameter $M_{10}$ in the direction of the Galactic centre; $M_{10}$ is the median $G$ magnitude (here in HEALPix regions of level 10) of \Gaia sources with \texttt{astrometric\_matched\_transits} $\leq 10$, reflecting the outcomes for faint sources of the \Gaia pipeline completeness decisions. The complex pattern results from the combination of the \Gaia scanning law and stellar density, which in turn depends on Galactic structure and dust distribution. Baade's window is prominent as a patch near $(\ell,b) = (2,-2)$, with bright $M_{10}$ presumably owing to the exceptionally high (true) source density. Several globular clusters are also visible, for instance the prominent M~22 near $(\ell,b)=(10,-7.5)$.
    }
    \label{fig:mapM10}
\end{figure*}

The best indicator of completeness we could identify is the $G$ magnitude of the sources with the smallest number of observations. The number of observations used to compute the astrometric solution of a given source is given in the \Gaia~DR3 catalogue as \texttt{astrometric\_matched\_transits}\footnote{The DR3 catalogue also contains the column \texttt{matched\_transits}, which counts all transits matched to a certain source even if they were not used in the construction of the catalogue. The quantity used in this study is \texttt{astrometric\_matched\_transits}.}. By construction, its minimum value is five because sources with fewer observations were not included. In the remainder of this paper, we denote $M_{10}$ the median magnitude of the sources with \texttt{astrometric\_matched\_transits}\,\,$\leq$ 10 in a given patch of sky. Its value is generally between 19 and 21.5 and strongly depends on how many times a given region was seen by \Gaia and how crowded the region is. This is illustrated in Fig.~\ref{fig:mapM10}, where the patterns introduced by stellar density and the \Gaia scanning law are clearly visible.

We choose the median value rather than the mean because some bright sources might occasionally have a small number of matched transits, and the median is a more robust summary statistic.
The model could also be calibrated on the median magnitude of sources with exactly five astrometric matched transits, but the chosen value of ten conveniently allows for a sufficient number of tracers even in sparse regions (the sparsest HEALPix level 7 contains 19 such sources), while keeping the total number of tracers manageable when building all-sky maps (210\,million in \Gaia~DR3).

The variation of completeness with the $M_{10}$ value of each investigated patch of sky is shown in Fig.~\ref{fig:c_vs_M10_magRanges}. The completeness in any magnitude range is overall a tight function of $M_{10}$, although a dispersion of up to $\sim0.1$ can be seen for some magnitude ranges. This dispersion effectively sets the limit of the precision one can achieve by using $M_{10}$ as the sole predictor. The same effect is illustrated in the bottom-left panel of Fig.~\ref{fig:obs_and_model}, where it can be seen that a given value of $M_{10}$ can correspond to slightly different completeness profiles. The effect of crowding likely depends not only on the true source density but also on the magnitude distribution of the sources, and a different parameter derived from the distributions shown in the top panel of Fig.~\ref{fig:amt_vs_Gmag} might be able to provide a second-order correction to the simple model presented in this study.

\subsection{Source count ratios relative to DECaPS}

We study the completeness in $1085$ patches of size $18 \times 7.2$\,arcmin across the DECaPS DR1 footprint, sampling a wide range of source densities. The size of the patches was chosen to allow us to avoid gaps in the coverage of the DECaPS DR1 data, mainly present near the Galactic centre. The distribution of those patches is shown in Fig.~\ref{fig:map_decaps_fields}.
Patches containing at least 10,000 \Gaia sources are further divided into up to eight bins to provide a finer spatial resolution in densest areas, for a total of 2,906 individual regions. We spatially match the \Gaia data to DECaPS with a 1\,arcsecond radius. When several DECaPS sources are present within this radius, we consider the best match to be the source whose $r$ magnitude is closest to the \Gaia source's $G$. Matches with a magnitude difference larger than 1\,mag are discarded. We find that less than 0.5\% of \Gaia sources have no DECaPS counterpart. The colour and magnitude of these missing sources seem to be a random subset of the \Gaia data. They appear to follow lines of constant declination on the sky, which suggests that they correspond to an instrumental effect of the Dark Energy Camera (e.g. bleeding trails caused by the presence of bright stars) rather than spurious \Gaia detections.
A small fraction of the \Gaia sources ($\sim$0.3\%, see Fig.~\ref{fig:map_noG}) lack a $G$-band magnitude. Since the present study investigates the completeness of the \Gaia catalogue as a function of magnitude, our procedure treats these sources as if they were missing from the \Gaia catalogue.

\begin{figure*}
    \centering
    \includegraphics[width=0.95\textwidth]{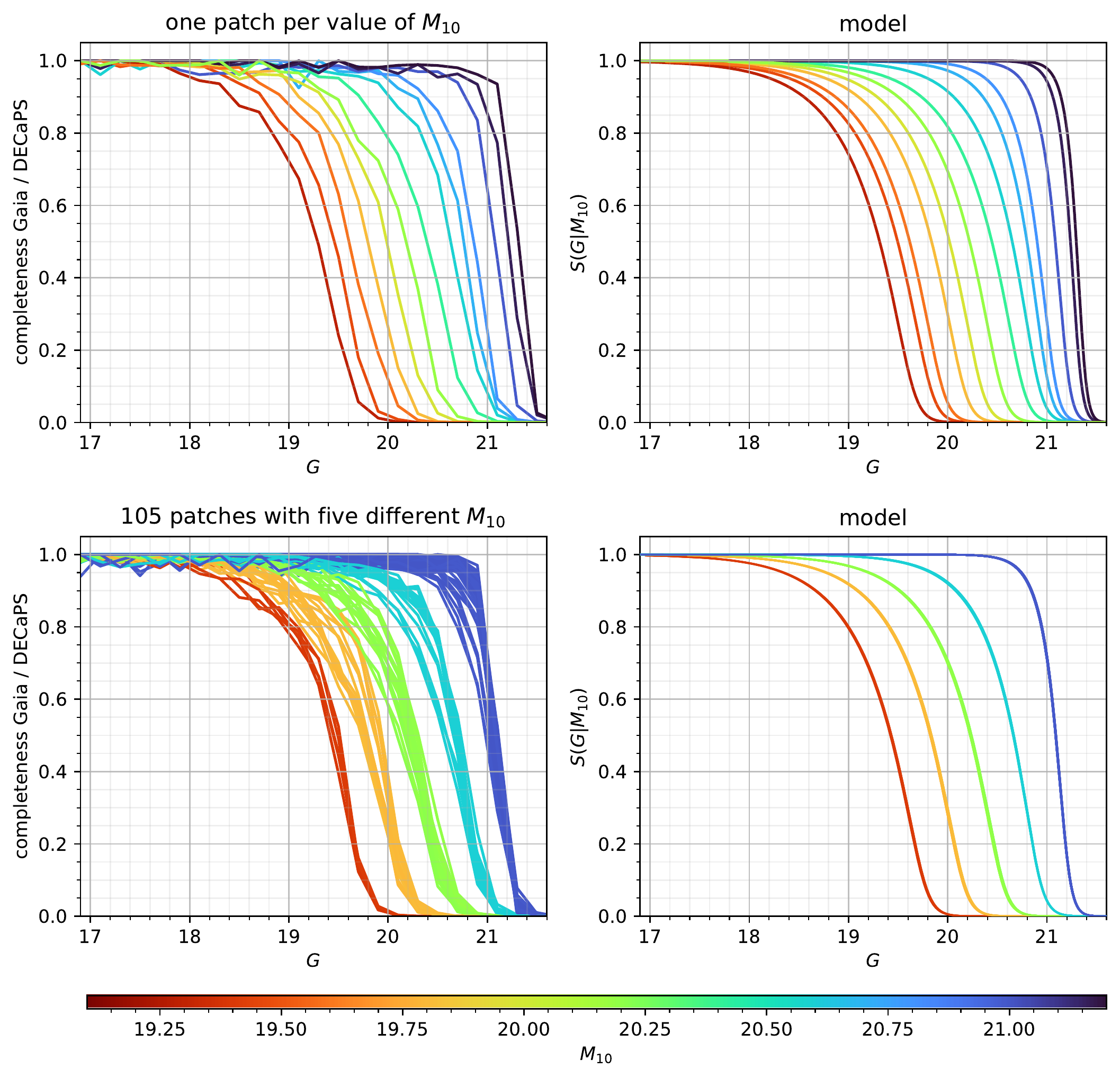}
    \caption{\Gaia's catalogue completeness $S(G~|~M_{10})$ as a function of $G$ magnitude for a given completeness parameter $M_{10}$. Top left: empirically determined \Gaia $G$ completeness derived from the comparison with DECaPS photometry in thirteen DECaPS patches that are each colour-coded by their $M_{10}$. Top right: model completeness, from Eq.~\ref{eq:sigmoid}, for the corresponding $M_{10}$ values. Bottom left: empirically-determined completeness derived from from the comparison with DECaPS photometry in 105 DECaPS patches with $M_{10}$=19.4, 19.8,20.2, 20.6, and 21 (within 0.01\,mag). The dispersion at a given $M_{10}$ corresponds to the scatter seen in Fig.~\ref{fig:c_vs_M10_magRanges} (right panel) and the residuals in Fig.~\ref{fig:residuals_tworanges}. Bottom right: model completeness for these five values of $M_{10}$.
    }
    \label{fig:obs_and_model}
\end{figure*}

We estimate the $G$ magnitude of the missing \Gaia sources from their DECaPS $(r,r-i)$ photometry. The conversion is performed by fitting a linear relation with the form $G = a*r + b*(r-i) + c$ in each patch of sky separately to account for the fact that photometric transformations are extinction-dependent. We then compute the fraction of DECaPS sources with a \Gaia counterpart in bins of $G$ magnitude of width 0.2\,mag, from $G=15$ to 23.
The completeness as a function of magnitude is shown in Fig.~\ref{fig:obs_and_model}, colour-coded by the value of $M_{10}$ for each region. In the densest regions, the completeness reaches 50\% at $G\sim19$, while in sparse regions that benefited from large numbers of observations, the \Gaia catalogue appears essentially complete down to $G\sim20.5$.

\subsection{Fitting the model} \label{sec:fitting}

We model the completeness curve computed in each region with a sigmoid function. To capture the change of slope from dense to sparse regions as well as the slight asymmetry of the curve, we define a \emph{generalised sigmoid} with the baroque but flexible analytic form:

\begin{equation}
    S(G~|~M_{10}) = 1 - 0.5 \times \left[ \tanh \left( \frac{x(M_{10}) - G}{y(M_{10})}  \right) + 1 \right] ^{~z(M_{10})}  ,  \label{eq:sigmoid}
\end{equation}

\noindent where $x$ is the magnitude of the inflexion point, $y$ controls how steeply the completeness drops at the inflexion point (smaller values correspond to a steep decrease), and $z$ describes the skewness of the function ($z<1$ means it is flatter at bright magnitudes). The effect of varying these three parameters is illustrated in Fig.~\ref{fig:sigmoid_model}.

We initially fit the triplet ($x,y,z$) independently in each of the 2,906 patches. The generalised sigmoid defined in equation (\ref{eq:sigmoid}) allows the fit to reproduce the observed completeness of all patches with residuals smaller than 2\%. Unfortunately, we cannot compute and provide an all-sky map of these three parameters because they can only be derived directly where DECaPS data is available. Instead, we investigate and model the relation between the three parameters and $M_{10}$.

The parameter $x$ (which roughly sets the magnitude of 50\% completeness) scales almost linearly with $M_{10}$. The parameters $y$ and $z$ (describing the slope and skewness of the curve) mostly follow two regimes, remaining roughly constant when $M_{10} < 20.5$, then increasing in value which results in a steeper, less asymmetric shape of the completeness curve for higher values of $M_{10}$. The high-$M_{10}$ regime corresponds to the flux-limited selection function, where transiting sources are only granted a detection window if the \Gaia sky mapper measures a magnitude $G_{onboard}<20.7$ for that particular transit \citep{gaiaMission}. The low-$M_{10}$ regime corresponds to areas on the sky where crowding plays a major role in the selection function.

To capture the variation of $x$, $y$, and $z$ with $M_{10}$ through these two regimes,
we model them as a broken slope relation, with the same location $M_{break}$ of the break for all three:

\begin{equation}
  x(M_{10})=\begin{cases}
    a_x M_{10} + b_x  & \text{if $M_{10} < M_{break}$}\\
    c_x M_{10} + (a_x - c_x ) M_{break} + b_x  & \text{otherwise}
  \end{cases}
\end{equation}

\begin{equation}
  y(M_{10})=\begin{cases}
    a_y M_{10} + b_y  & \text{if $M_{10} < M_{break}$}\\
    c_y M_{10} + (a_y - c_y ) M_{break} + b_y  & \text{otherwise}
  \end{cases}
\end{equation}

\begin{equation}
  z(M_{10})=\begin{cases}
    a_z M_{10} + b_z  & \text{if $M_{10} < M_{break}$}\\
    c_z M_{10} + (a_z - c_z ) M_{break} + b_z  & \text{otherwise}
  \end{cases}
\end{equation}

\begin{figure*}
    \centering
    \includegraphics[width=0.95\textwidth]{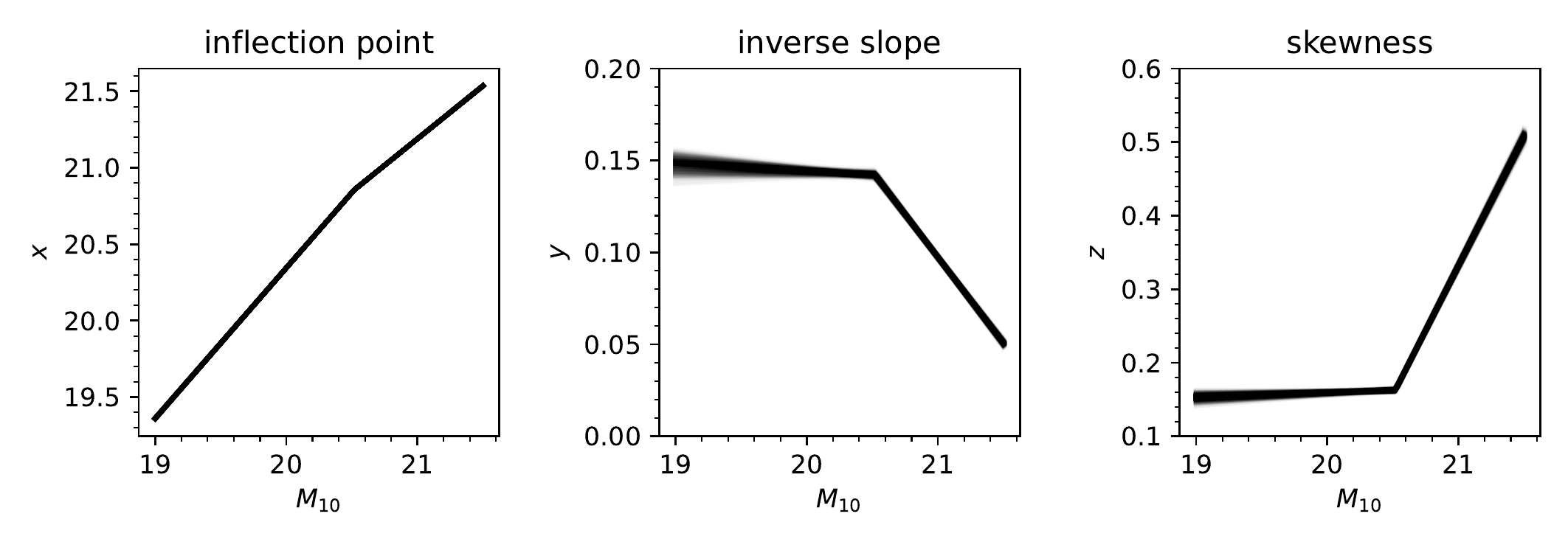}
    \caption{ Relation between the parameters of the sigmoid and $M_{10}$ in Eq.~\ref{eq:sigmoid}, for 500 samples from the MCMC chain: the inflection point, $x(M_{10})$ -- Eq.2, on the left; the inverse slope, $y(M_{10})$ -- Eq.3, at center; and the skewness,$z(M_{10})$ -- Eq.4, on the right.   The magnitude of 50\% completeness, $x(M_{10})$ is well approximated by $M_{10}$. Fields with fainter $M_{10}$ have a steeper (smaller $y(M_{10})$ ) and more symmetrical ($z(M_{10})$ towards 1) selection function. Examples of resulting sigmoids are shown in Fig.~\ref{fig:obs_and_model} for a range of $M_{10}$ values.
    }
    \label{fig:emcee_hyperParams_vs_M10}
\end{figure*}

The resulting hierarchical model has ten free hyperparameters (three for each of $x$, $y$, and $z$, plus the location of the break). We add a final free parameter $\sigma$ representing the noise on the observed completeness profiles.
The noise is assumed to be Gaussian and constant with magnitude. This is a rough approximation, and here $\sigma$ acts like a nuisance parameter rather than a model for the noise. The corresponding log-likelihood is:

\begin{equation}
    -n \log \sigma - \frac{1}{2 \sigma^2} \sum_{i=1}^{n} (obs_i - pred_i)^2
\end{equation}

\noindent where $n$ is the total number of data points: the source count ratios in 40 magnitude bins $\times$ 2,906 patches. We maximise the log-likelihood with the Markov chain Monte-Carlo sampler \texttt{emcee} \citep{ForemanMackey13} and explore the parameter space with 32 walkers for 10,000 steps each.
We impose that $a_x$, $c_x$, and $c_z$ must be positive, $c_y$ must be negative, $\sigma$ must be between 0 and 1, and $M_{break}$ between 19 and 21. The priors on the other parameters are left unbounded.
The sampling takes about two hours on an 8-core laptop. We discard the first 1,000 iterations as burn-in\footnote{The resulting chain is 70 to 90 times longer (depending on the parameter) than the autocorrelation time estimated by \texttt{emcee}.}. 
We provide the median  of posterior samples for each parameter in Table~\ref{tab:mcmc}. The final relation between the parameters of the sigmoid and $M_{10}$ are shown in Fig.~\ref{fig:emcee_hyperParams_vs_M10}.

\begin{table}
\begin{center}
	\caption{ \label{tab:mcmc} Parameters for the selection function $S(G~|~M_{10})$ in Eq.~\ref{eq:sigmoid}.}

	\begin{tabular}{ c  c  c  c }
	\hline
	\hline

    parameter & median & \multicolumn{2}{c}{uncertainty} \\
    \hline
$a_x$  &  0.985  &  +0.002 & -0.001 \\
$b_x$  &  0.649  &  +0.030 & -0.031 \\
$c_x$  &  0.693  &  +0.001 & -0.002 \\
    \hline
$a_y$  &  -0.004  &  +0.003 & -0.003 \\
$b_y$  &  0.223  &  +0.060 & -0.051 \\
$c_y$  &  -0.093  &  +0.002 & -0.002 \\
    \hline
$a_z$  &  0.006  &  +0.003 & -0.003 \\
$b_z$  &  0.034  &  +0.068 & -0.060 \\
$c_z$  &  0.351  &  +0.004 & -0.004 \\
    \hline
$M_{break}$  &  20.519  &  +0.002 & -0.001 \\
\hline
$\sigma$  &  0.02060  &  +0.00004 & -0.00004 \\
	\hline
	\hline
	\end{tabular}
\tablefoot{The parameters for Eq.~\ref{eq:sigmoid} are defined via Eqs.2-4, and were obtained via MCMC. We adopt the median of the posterior chain as the best value, and state the 16$^{th}$ to 84$^{th}$ percentile confidence interval. Here, $\sigma$ is a nuisance parameter in the fitting procedure. The corner plot of the full posterior chain is shown in Fig.~\ref{fig:cornerplot_11}.  }
\end{center}
\end{table}

\begin{figure*}
    \centering
    \includegraphics[width=0.95\textwidth]{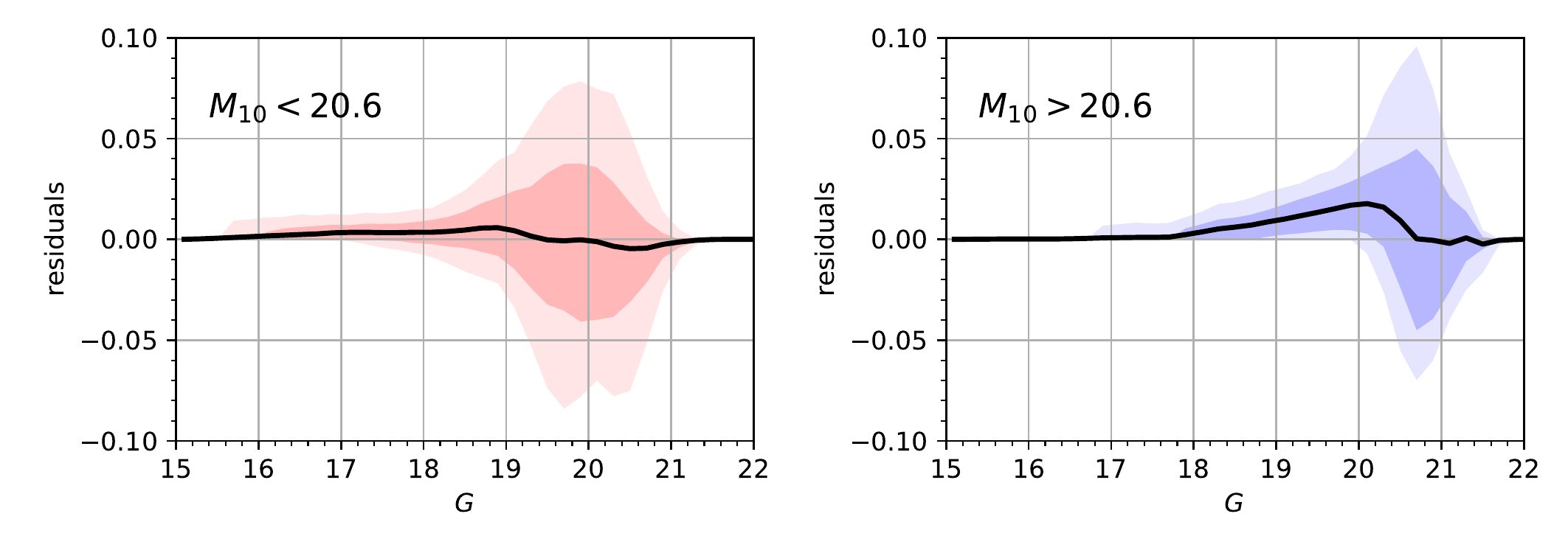}
    \caption{ Mean completeness residuals, i.e. model predicted minus observed completeness,  as a function of magnitude, for patches in two different ranges of $M_{10}$. The shaded areas correspond to the 16th to 84th percentile and 5th to 95th percentile intervals. Expectedly, the residuals are highest near 50\% completeness and smaller in the highly complete or dramatically incomplete regime. 
    }
    \label{fig:residuals_tworanges}
\end{figure*}

Figure~\ref{fig:residuals_tworanges} shows the mean and dispersion of the residuals, for two ranges of $M_{10}$. The prediction is most precise where the model predicts completeness of 0 or 100\%. Where the prediction is least precise, the dispersion of the residuals reaches about 5\%.

\begin{figure}
    \centering
    \includegraphics[width=0.49\textwidth]{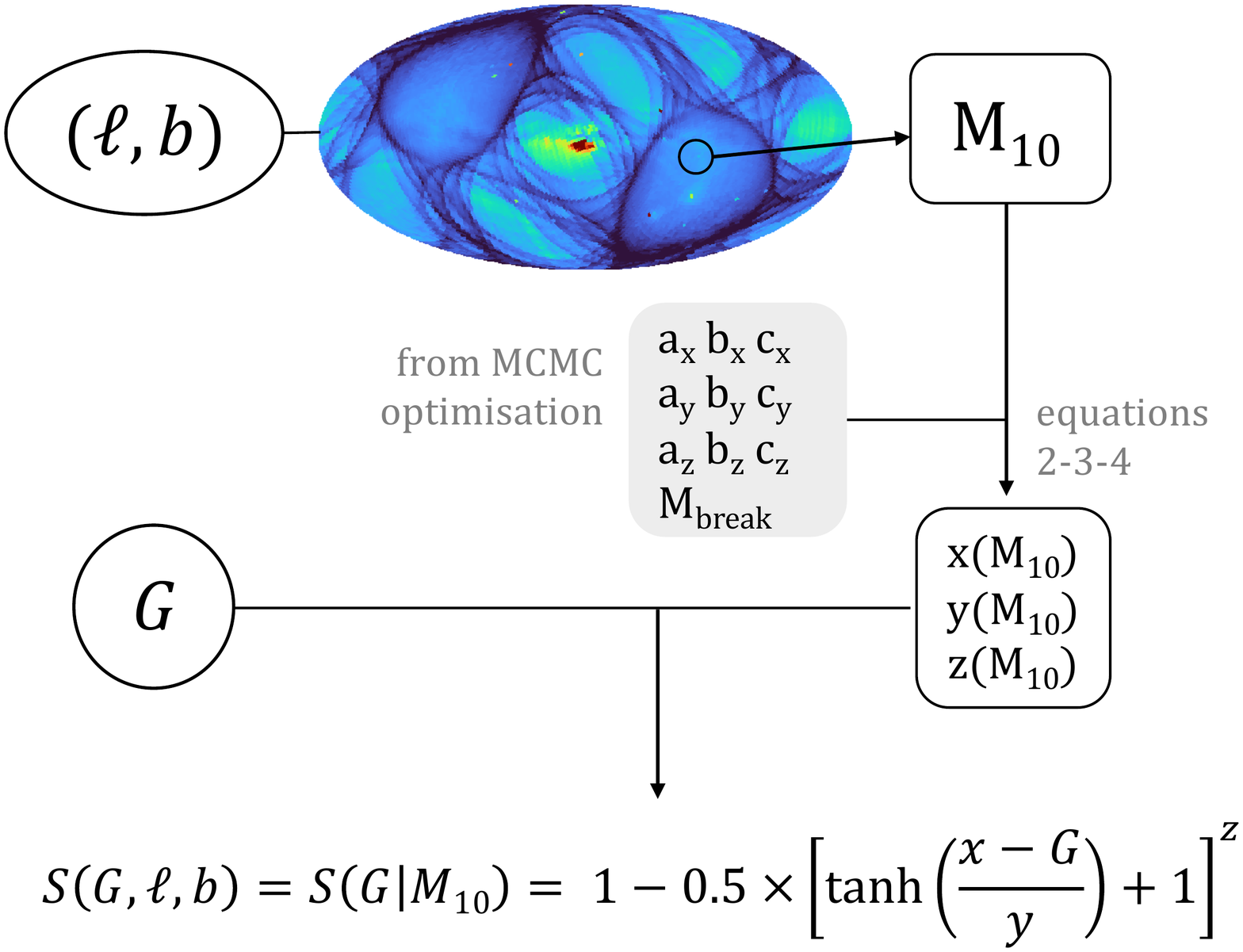}
    \caption{ Summary of the workflow used to build the selection function $S(G~|~M_{10})$ (Eq.~\ref{eq:sigmoid}) as a function of sky position and magnitude $G$.
    }
    \label{fig:workflow}
\end{figure}

Figure~\ref{fig:workflow} summarises the workflow and how we use the hyperparameters to predict the completeness as a magnitude function at any sky position.

\section{Testing the model} \label{sec:testing}

\subsection{With more DECaPS data}

\begin{figure*}
    \centering
    \includegraphics[width=0.85\textwidth]{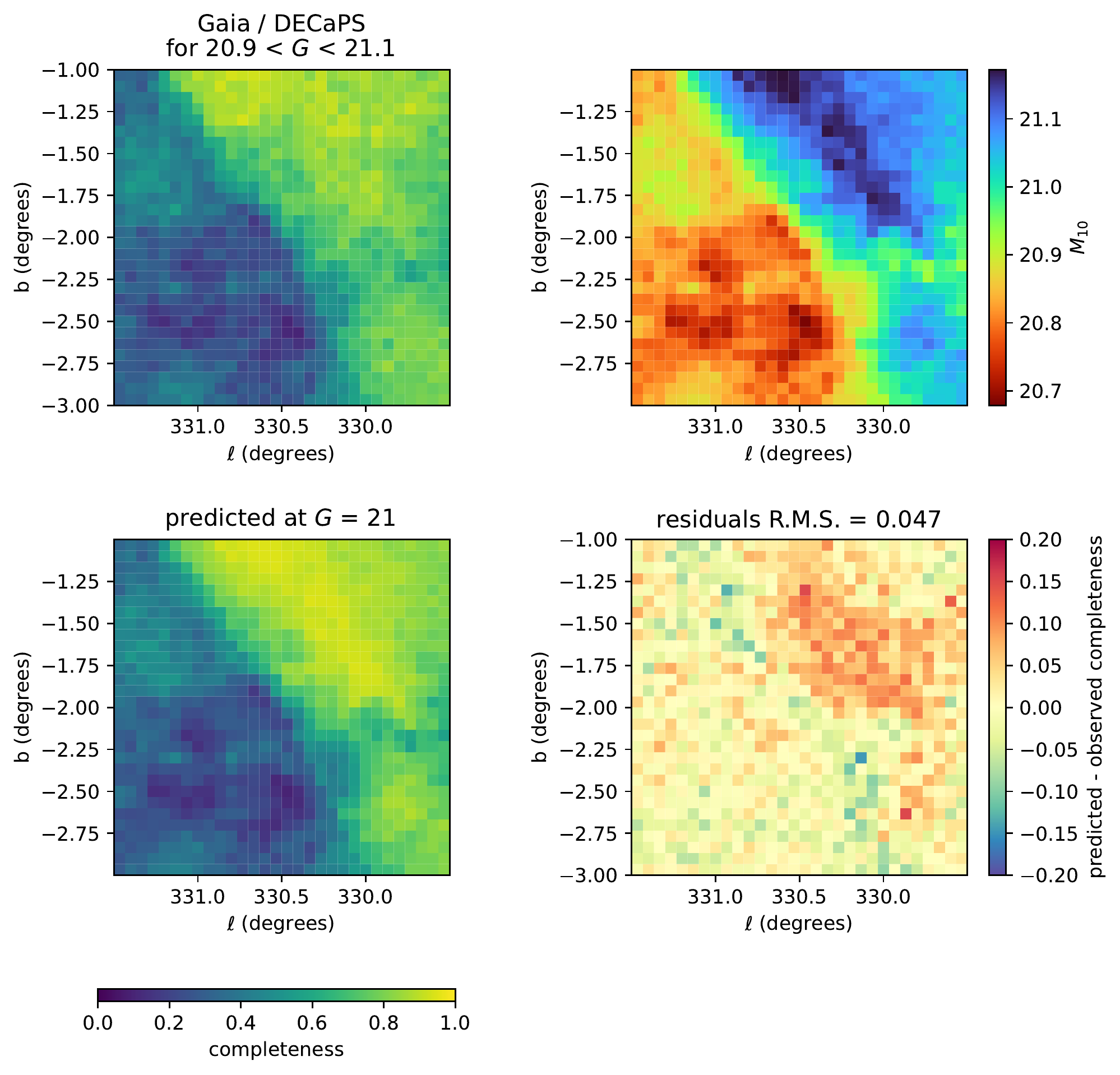}
    \caption{ Comparison of the direct, empirical and model-predicted completeness maps, illustrated at $G\sim 21$. Top left: map of the direct completeness estimate, i.e. the ratio of source densities in \Gaia and DECaPS in the magnitude range 20.9 < $G$ < 21.1. Top right: map of the quantity $M_{10}$ used to predict the model completeness. Bottom left: completeness at $G=21$ predicted from the $M_{10}$ map and the model of Eq.~\ref{eq:sigmoid}. Bottom right: map of the difference between the predicted and observed completenesses. Note that using more external information, the model-predicted completeness map (bottom left) is effectively a de-noised version of the empirical completeness map (top left).
    }
    \label{fig:map_prediction_330}
\end{figure*}

We verify the prediction of our model by applying it to regions of the DECaPS footprint that were not used for the fitting step. The $2^{\circ} \times 2^{\circ}$field of view shown in Fig.~\ref{fig:map_prediction_330} was chosen to straddle the boundary between a stripe which received a large number of visits and an adjacent area with much fewer transits. The $M_{10}$ parameter was mapped by computing the median $G$ magnitude of stars with ten or fewer \texttt{astrometric\_matched\_transits} in spatial bins of $2.4 \times 2.4$\,arcmin. The main diagonal feature, splitting the field of view in two, is due to the \Gaia scanning law. The finer structure is shaped by patchy dust extinction.

The model-predicted completeness map (bottom left of Fig.~\ref{fig:map_prediction_330}) obtained from the $M_{10}$ map (top right) is less noisy than the map obtained directly from source count ratios (top left). 
However, as discussed in Sect.~\ref{sec:improvements}, the use of $M_{10}$ as the sole predictor of completeness can lead to local biases of a few per cent (within the amplitude of the residuals shown in Fig.~\ref{fig:residuals_tworanges} and the bottom left panel of Fig.~\ref{fig:obs_and_model}). In Fig.~\ref{fig:map_prediction_330} this leads to slightly overestimating the completeness of the most complete area (with $M_{10}$>21.1).

\subsection{HST data of globular clusters} \label{sec:HST}

The cores of globular clusters (GCs) are among the most challenging regions for \Gaia due to their high densities. In some particularly dense clusters, the completeness at $G$=18 is close to zero.

\begin{figure}
    \centering
    \includegraphics[width=0.49\textwidth]{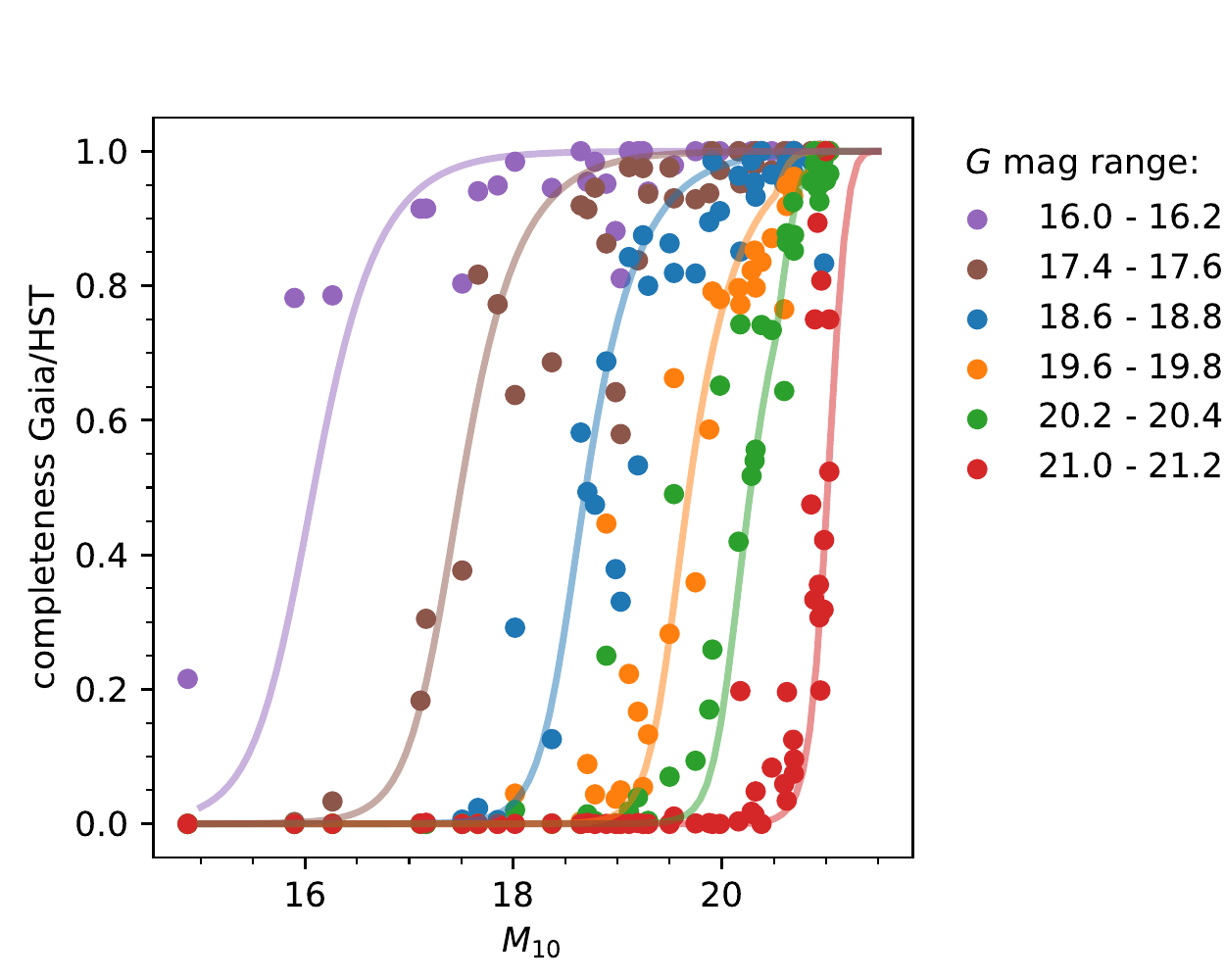}
    \caption{ Completeness of \Gaia relative to HST as a function of $M_{10}$ in five chosen magnitude ranges, for an extreme crowding regime: the core and outskirts of 26 globular clusters. The lines show the expected completeness according to our model. Remarkably, our model (Eq.~\ref{eq:sigmoid}) provides unbiased completeness predictions based on \Gaia information alone. }
    \label{fig:c_vs_M10_GCs}
\end{figure}

The data (presented in Sect.~\ref{sec:data}) contains observations of the inner 3\,arcmin of 26 GCs. We split each field of view into a core (inner 1.5\,arcmin) and a surrounding area. Since these objects are very dense and the field of view is smaller than the spatial binning we used in our model calibration, fifteen cluster regions have values of $M_{10}$ smaller than the range (19.11 to 21.23) covered by the calibration set.
We show in Fig.~\ref{fig:c_vs_M10_GCs} that our model can predict the completeness even when extrapolated to these crowded, low-$M_{10}$ fields. The extrapolation only seems to fail by 20\% in the most extreme case of crowding, which corresponds to the inner 1.5\,arcmin of Omega~Centauri ($M_{10}$)

\begin{figure*}
    \centering
    \includegraphics[width=0.85\textwidth]{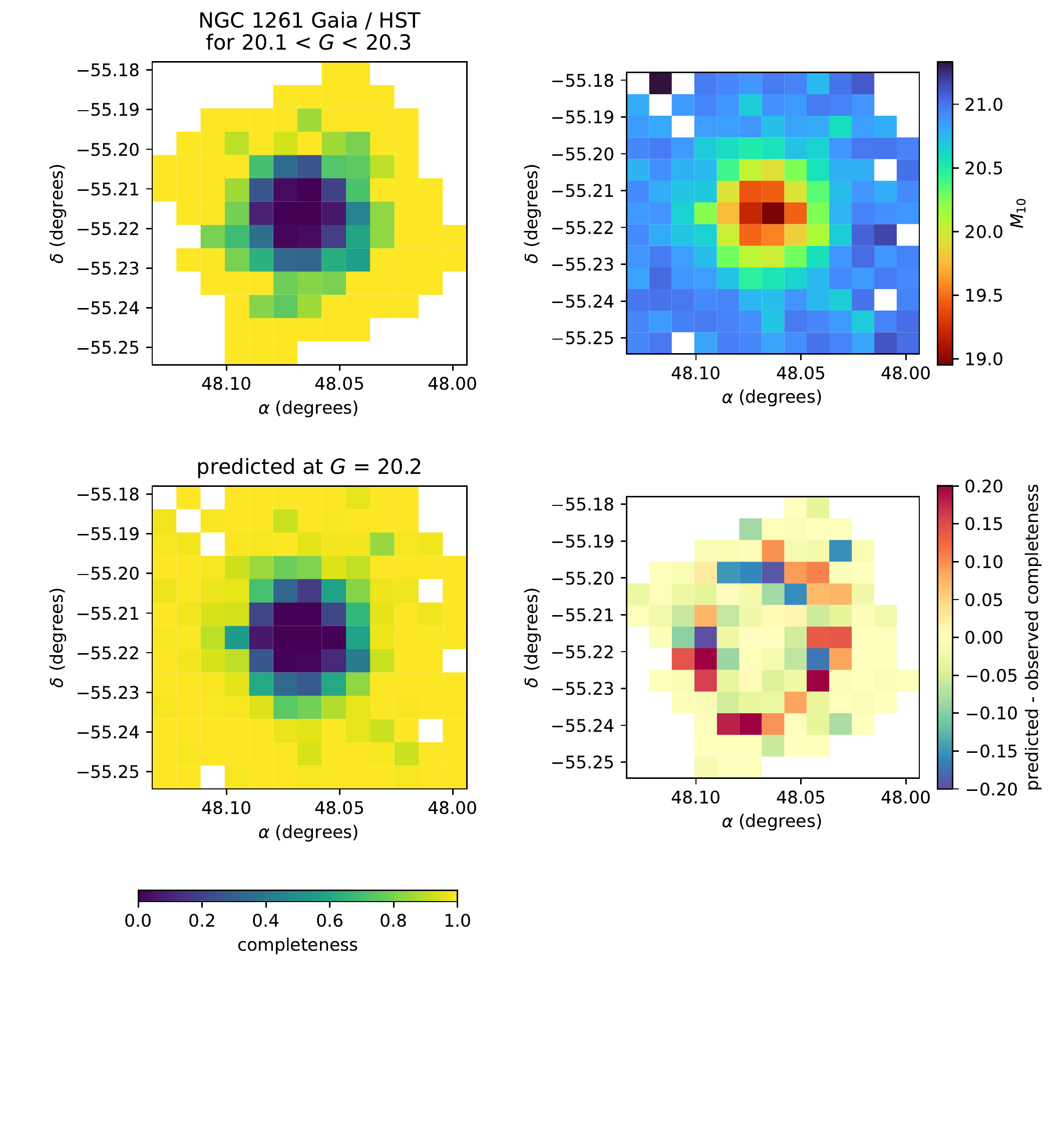}
    \caption{ 
    Comparison of the direct, empirical and model-predicted completeness maps, illustrated for the globular cluster NGC~1261. Top left: map ratio of the number of sources in \Gaia and HST 
    in the magnitude range 20.9 < $G$ < 21.1. Top left: map of $M_{10}$ used to predict the completeness. Bottom left: predicted completeness at $G=21$. Bottom right: map of the difference between the predicted and observed completeness: the variance is largest in the intermediate completeness regime (see Fig.~\ref{fig:residuals_tworanges}), producing a ring-like structure in the residuals map.
    }
    \label{fig:map_ngc_1261}
\end{figure*}

As an example, we map the completeness prediction of globular cluster NGC~1261 in Fig.~\ref{fig:map_ngc_1261}. Our model correctly identifies the regions of 100\% and 0\% completeness. The intermediate regions appear as a ring of noise on the residuals map (bottom-right panel of Fig.~\ref{fig:map_ngc_1261}), due to the small spatial binning and the narrow magnitude range.

\subsection{Comparison to the Gaiaverse model} \label{sec:gv}

\begin{figure*}
    \centering
    \includegraphics[width=0.99\textwidth]{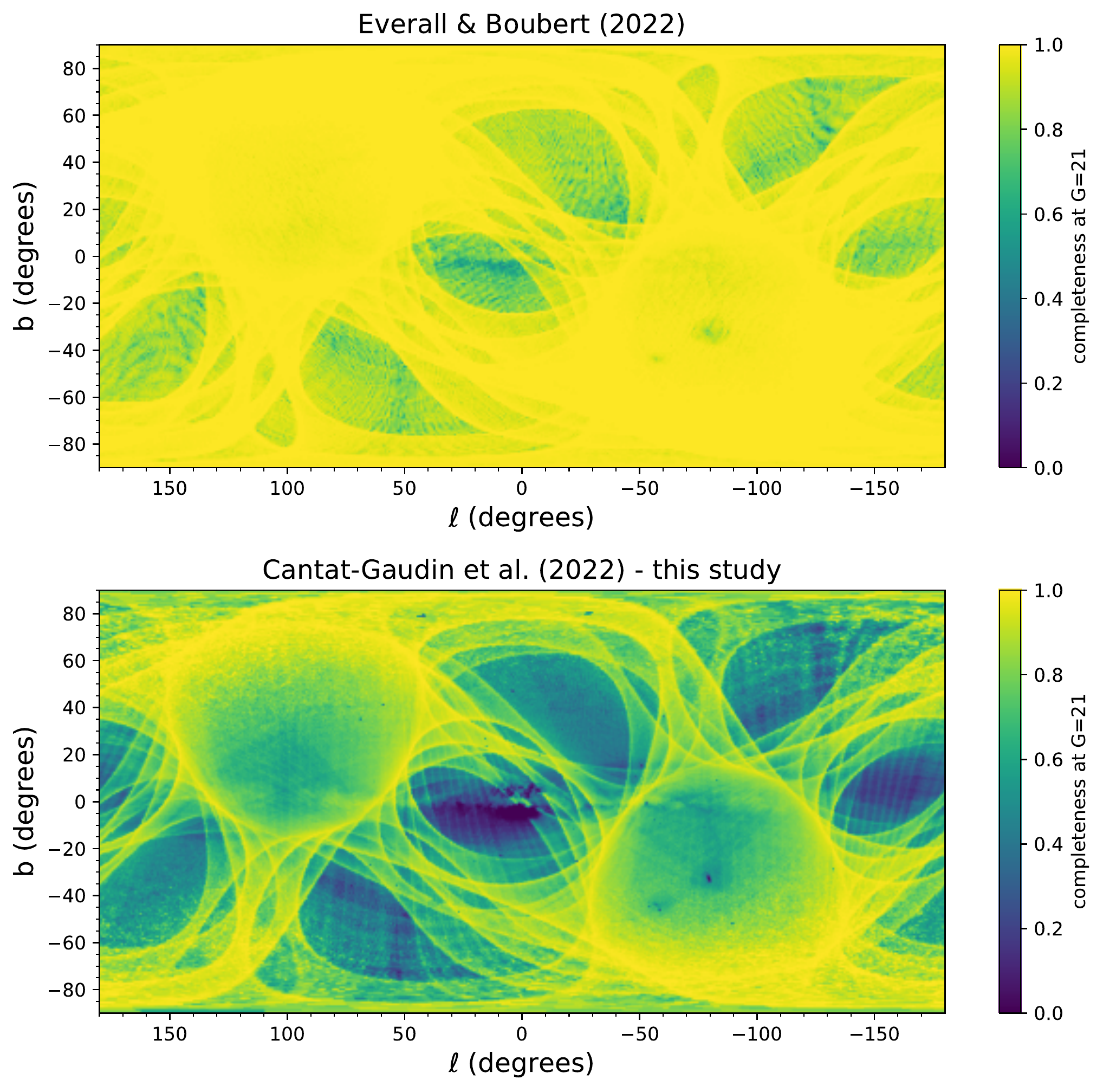}
    \caption{ Global comparison of the completeness maps predicted by the \emph{ab initio} completeness model \citep[][top panel]{Everall22gaiaverse5}, and our empirically derived $M_{10}$-based completeness model (bottom). The overall morphology of the two maps is similar, but our empirical completeness model implies far greater incompleteness (at $G$=21), especially in the regions of high source densities.
    }
    \label{fig:comparisons_GV_21}
\end{figure*}

We compare our predictions with the model of \citet{Everall22gaiaverse5} (hereafter EB22), which itself is the \Gaia DR3 update of the model developed by \citet{Boubert20gaiaverse2} for DR2. Their predictions do not rely on comparisons to reference data but on a model of the \Gaia scanning law and of the detection efficiency as a function of magnitude.
We show the all-sky map of completeness at $G$=21 predicted by both models in Fig.~\ref{fig:comparisons_GV_21}.

\citet{Boubert20gaiaverse2} point out that modelling the effects of crowding is a complex task, as one can only know the observed density of sources while crowding depends on the true density (including those missing from the catalogue). For this reason, the effect of globular clusters or high-density regions near the Galactic centre is more clearly visible in our model, which naturally accounts for crowding via the $M_{10}$ parameter.

The most striking difference between both models is that EB22 predict much higher completeness even in non-crowded regions, with an essentially 100\% complete catalogue at $G$=21 across most of the sky, while our model predicts that a 100\% completeness at this magnitude is only achieved in the regions most favoured by the scanning law. This is supported by comparisons to the DECaPS data (including those used to calibrate the model), which show most of the Galactic plane is only 60-80\% complete at this magnitude.

A likely explanation for this discrepancy is that EB22 overestimates faint source detection probabilities. Figure~7 from EB22 shows that for sources with $G$=21, the reconstructed detection probability is $\sim$30\%, which translates to a 90\% probability of having 5 detections after 25 scans (the number for the least visited regions in DR3) and 99.5\% probability after 40 scans (the median number for the whole sky). The EB22 model estimates detection efficiencies from photometric time series published in the \Gaia archive for variable stars. This sample is likely to be biased towards stars with high-quality measurements and low photometric errors. A more realistic estimate of the range of detection efficiencies in a given region can be obtained from individual transit data of all \Gaia sources. This data is not publicly available and will be used in a future publication (Castro-Ginard et al., in prep.)

\section{Discussion} \label{sec:discussion}

\subsection{Limits and potential improvements} \label{sec:improvements}

With its three parameters, the generalised sigmoid functional form defined in Equation~\ref{eq:sigmoid} turns out to be sufficiently flexible to approximate the observed \Gaia-to-DECaPS count ratio as a function of magnitude in any region to within two per cent. Given the limited sky coverage of our ground truth catalogue (DECaPS, which only covers $\sim$7\% of the celestial sphere), we need to predict the values of Eq.~\ref{eq:sigmoid}'s three parameters, using the quantity $M_{10}$ defined in Sect.~\ref{sec:dependencies} and computed from the \Gaia data itself. The scatter observed in the right panel of Fig.~\ref{fig:c_vs_M10_magRanges} varies with $M_{10}$ and with $G$ magnitude and illustrates the limitations of predicting selection function $S(G~|~M_{10})$ using $M_{10}$ as the sole predictor of completeness. 

We could not identify a single quantity providing a better precision than $M_{10}$, which encodes the combined effect of the scanning law and crowding. Nonetheless, it may be possible to establish second-order corrections based on other \Gaia-derived quantities. The detection probability as a function of magnitude likely depends on the magnitude distribution of sources in a given field of view, not just on their total number. We investigated the residuals of our model but only found hints of additional correlations between completeness and the total number of scans, or completeness and observed source density, in some restricted ranges of $M_{10}$ and $G$ magnitude. 
Establishing the right functional form and choice of dependency for such ad-hoc corrections would be a difficult task. One might be tempted to follow a machine-learning approach and let the machine determine the most relevant predictors of the \Gaia-to-DECaPS count ratio. This would, however, incur the likely risk of overfitting, as many of the correlations in the reference data cannot reliably be generalised to the entire sky unless they can be supported by some understanding of the instrumental pipeline.  

A potential improvement to the $M_{10}$ proxy could be to characterise the entire distribution of \texttt{astrometric\_matched\_transits} with magnitude $G$ (illustrated in Fig.~\ref{fig:amt_vs_Gmag} for five chosen patches) rather than just its value at the faint end. The slope and shape of the drop in the number of matched transits may contain information on the level and type of crowding affecting the observations.
Another direction of improvement to explore is to see whether sky areas with a broader range of scanning angles are more likely to be complete since the sources missed by \Gaia are more likely to be different at each visit \citep{gaiaMission, Pancino17}. The dispersion in scanning angles over a given area could therefore be an additional parameter in the empirical description of the selection function. This quantity is available for each source in the \Gaia catalogue as \texttt{scan\_direction\_strength\_k2}, but testing its validity as a secondary predictor of completeness would be difficult in the present context because the regions with the densest clustering of scanning angles are located outside the DECaPS footprint \citep[see e.g. Fig. 1a in][]{Everall21gaiaverse4}. Using the DECaPS DR2 release \citep{Saydjari22}, which more than doubles the survey area, could mitigate this problem in future work.

Finally, the model constructed in this study assumes that the ($x$,$y$,$z$) parameters of the sigmoid are related to $M_{10}$ via a broken-slope relation, with a total of ten free hyperparameters. A more complex model (for instance, with more breaks) would decrease the residuals shown in Fig.~\ref{fig:residuals_tworanges} (and smooth out the kink near $G\sim20.2$ in its right panel), but unless the increase in complexity can be justified by some knowledge of the instrumental behaviour of \Gaia, a simpler model is more likely to be valid outside the DECaPS footprint.

\subsection{Dust extinction makes \Gaia more complete}

A natural but perhaps counter-intuitive effect of interstellar extinction is to increase completeness as a function of apparent magnitude. Foreground sources of a given apparent magnitude $G$ are more easily detected when projected against a ``dark'' background.

Of course, dust extinction still reduces the probability of \Gaia catalogue membership for sources of a given set of \emph{physical} properties and distance. Given that the selection function must be phrased in terms of \emph{observables}, modelling sources as a function of distance and absolute magnitude requires a 3D extinction map.

\subsection{Arguments of the selection function}

In this study, we expressed the \Gaia source catalogue selection function as a function of magnitude and position $(G,\ell,b)$. We find no evidence that this fundamental \Gaia selection function depends significantly on the source colour: in a given part of the sky, two sources with the same $G$ magnitude appear to have equal probabilities to be included in \Gaia, regardless of their colour. This result is not surprising, because the \Gaia sky mapper and the astrometric instruments on board the spacecraft operate in the $G$ band.

We point out that due to strong correlations between the observables, investigating the chromaticity of the selection function is a much more complex task than simply expressing detection rates as a function of colour. For astrophysical reasons, red stars tend to be intrinsically fainter than blue stars. Interstellar extinction acts in the same direction, making sources appear both fainter and redder. On the other hand, areas of the sky heavily obscured by dust are redder but also more complete due to the background being less crowded in the magnitude range where \Gaia operates.

\subsection{Selection function for subsets of the \Gaia catalogue}

This paper only addresses the completeness of the sample of \Gaia catalogue entries with a published position and $G$ magnitude, establishing the selection function noted $S_{\cat}^{\mathrm{parent}}(G,\ell,b)$ in the notation of \citet{Rix21}. In practice, most users will be interested in comparing other \Gaia quantities with theoretical models, such as observed $G_{BP}$ and $G_{RP}$ fluxes, parallaxes, proper motions, or more advanced data products provided by the \Gaia pipelines such as astrophysical parameters \citep{Creevey22cu8}. It is not clear whether the approach used in this study is suitable for selecting further, more restricted subsets of the \Gaia data. 

First, $M_{10}$ might not be a good predictor of the completeness of \Gaia subsets, say stars with spectra from the radial velocity spectrometer \citep{gaiaMission} (RVS),  because different instruments on board the spacecraft have different crowding limits: 1,050,100 sources per square degree for the astrometric instrument, 750,000 for the BP/RP spectrographs, and 35,000 for RVS. It may, however, be possible to construct equivalent quantities to characterise particular subsets, e.g. an equivalent of $M_{10}$ for Gaia sources with BP/RP. Second, it is not clear that the generalised sigmoid function (Eq.~\ref{eq:sigmoid}) is a good functional form for the selection functions of various \Gaia subsets. Third, the selection function of some subsets will depend on more than just $G$ magnitude and sky position. For instance, \citet{Everall20seestar} and \citet{Rybizki21rvs} express the RVS and $\texttt{ruwe}<1.4$ completeness as functions of ($G-G_{RP}$).

For thinking about the construction of more complex selection functions, we refer the reader to \citet{Rix21}, who provide recommendations on how to construct the sample function $S_{\cat}^{\mathrm{sample}}(\vq)$ of a given subset of the \Gaia data selected on attributes $\vq$. In general, an overall selection function can be approximated as  a multiplication reflecting the different Boolean steps in the sample selection:

\begin{equation}
    S_{\cat}(\vq,G,\ell,b) = S_{\cat}^{\mathrm{sample}}(\vq) \times  S_{\cat}^{\mathrm{parent}}(G,\ell,b)
\end{equation}

\section{Summary and conclusion} \label{sec:conclusion}

This study is part of a paper series by the GaiaUnlimited project that aims to characterise the \Gaia selection function and provide the astronomical community with corresponding data and tools.
This paper presents an analytical model of the \Gaia DR3 completeness as a function of observed $G$ magnitude and position on the sky. Our model depends on a single quantity which is derived from the \Gaia data itself: the median magnitude $M_{10}$ in a patch of the sky of catalogued sources with \texttt{astrometric\_matched\_transits}\,\,$\leq 10$. $M_{10}$ reflects the elementary processes and decisions made by the \Gaia pipeline to turn observations into the published \Gaia catalogue and naturally accounts for the effects of crowding and the \Gaia scanning law.

As ground truth, we rely on the DECaPS survey, which is deeper than \Gaia and whose pipeline is optimised for high-density fields, to calibrate our model. We test our predictions against DECaPS and Hubble Space Telescope observations of globular clusters. Our model predicts the observed completeness with a precision of up to a few per cent. We make the model available as a Python package through the GaiaUnlimited web page, along with documentation and tutorials.

The present model only provides a selection function for the \Gaia DR3 entries with a published $G$ magnitude and sky position. Sub-samples of this catalogue will be characterised in upcoming GaiaUnlimited publications.

\section*{Acknowledgments}

This work is a result of the GaiaUnlimited project, which has received
funding from the European Union's Horizon 2020 research and innovation program
under grant agreement No 101004110. The GaiaUnlimited project was started at
the 2019 Santa Barbara Gaia Sprint, hosted by the Kavli Institute for
Theoretical Physics at the University of California, Santa Barbara.

This work has made use of data from the European Space Agency (ESA) mission
{\it Gaia} (\url{https://www.cosmos.esa.int/gaia}), processed by the {\it
Gaia} Data Processing and Analysis Consortium (DPAC,
\url{https://www.cosmos.esa.int/web/gaia/dpac/consortium}). Funding for the
DPAC has been provided by national institutions, in particular, the
institutions participating in the {\it Gaia} Multilateral Agreement.

A.~R.~C. is supported in part by the Australian Research Council through a Discovery Early Career Researcher Award (DE190100656) and through Discovery Project DP210100018.

DECaPS data were retrieved with the \texttt{astro-datalab}\footnote{\url{https://github.com/astro-datalab/datalab/}} Python package, and the \Gaia data with \texttt{astroquery} \citep{2019AJ....157...98G}. This work also made use of the Python packages \texttt{astropy} \citep{2018AJ....156..123A}, \texttt{scipy} \citep{2020SciPy-NMeth}, \texttt{astroML} \citep{astroML}, \texttt{MWplot}\footnote{\url{https://pypi.org/project/mw-plot/}}, \texttt{numpy} \citep{harris2020array}, \texttt{plotly} \citep{plotly}, \texttt{healpy}\footnote{\url{http://healpix.sourceforge.net}} \citep{2005ApJ...622..759G,Zonca2019}, \texttt{pandas} \citep{mckinney2010data}, and \texttt{matplotlib} \citep{Hunter:2007}. TCG acknowledges an extensive use of TOPCAT \citep{Taylor05} and Jupyter notebooks \citep{Kluyver2016jupyter}.

\bibliographystyle{aa}
\bibliography{refs}

\appendix

\section{Observed \Gaia source density and true source density}
\begin{figure*}
    \centering
    \includegraphics[width=0.99\textwidth]{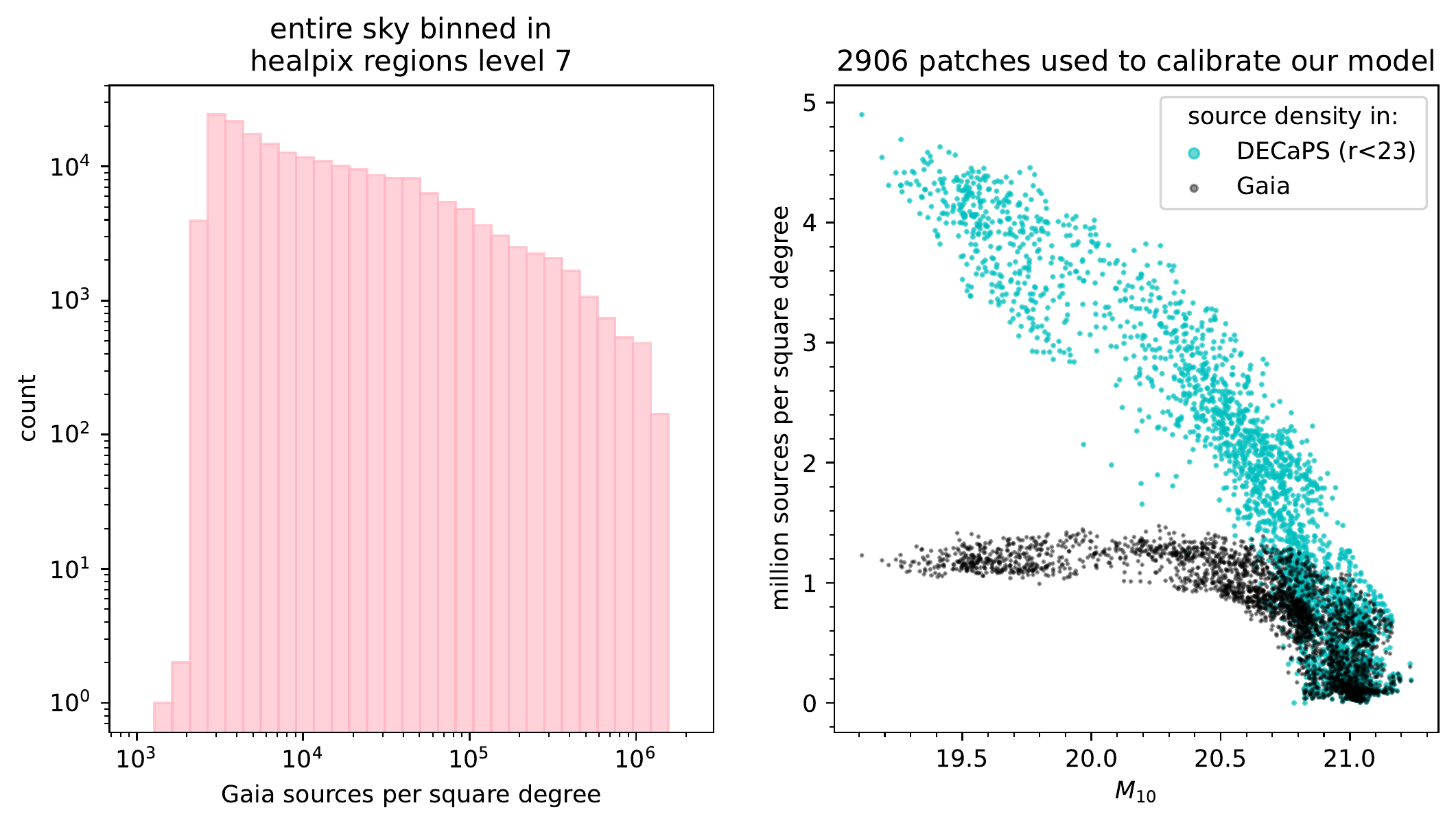}
    \caption{  Left: distribution of \Gaia~DR3 sources densities computed in 196,608 HEALPix regions level 7 across the whole sky. Crowding artificially sets an upper limit of the catalogue source density. Right: source density as a function of $M_{10}$ in \Gaia~DR3 (black) and DECaPS with $r$<23 (cyan) in the 2906 patches we use to calibrate our model (see Fig.~\ref{fig:map_decaps_fields}). Outside the sparsest regions, the \Gaia source density provides almost no information on the true source density and level of crowding.
    }
    \label{fig:gaia_density_saturation}
\end{figure*}

The crowding limit of the astrometric instrument on board the \Gaia spacecraft is 1,050,000 sources per square degrees \citep{gaiaMission}. In denser regions, the processing pipeline is not able to track all sources transiting through the focal plane and prioritises bright sources. 
Since the set of discarded faint stars can be different every time \Gaia scans a given region, the final source density in the \Gaia catalogue locally reaches values as high as $\sim$1.4\,million sources per square degree, but the crowding issue artificially sets an upper limit on the catalogue source density, as shown in the left panel of Fig.~\ref{fig:gaia_density_saturation}.

The right panel of Fig.~\ref{fig:gaia_density_saturation} shows that the parameter $M_{10}$ scales almost linearly with true source density (approximated by the density of DECaPS sources brighter than $r$=23), while the \Gaia source density is essentially flat for $M_{10}$<20.5 and never reaches above 1.5\,million sources per square degree.

\section{Number of matched transits with magnitude}
\begin{figure*}
    \centering
    \includegraphics[width=0.95\textwidth]{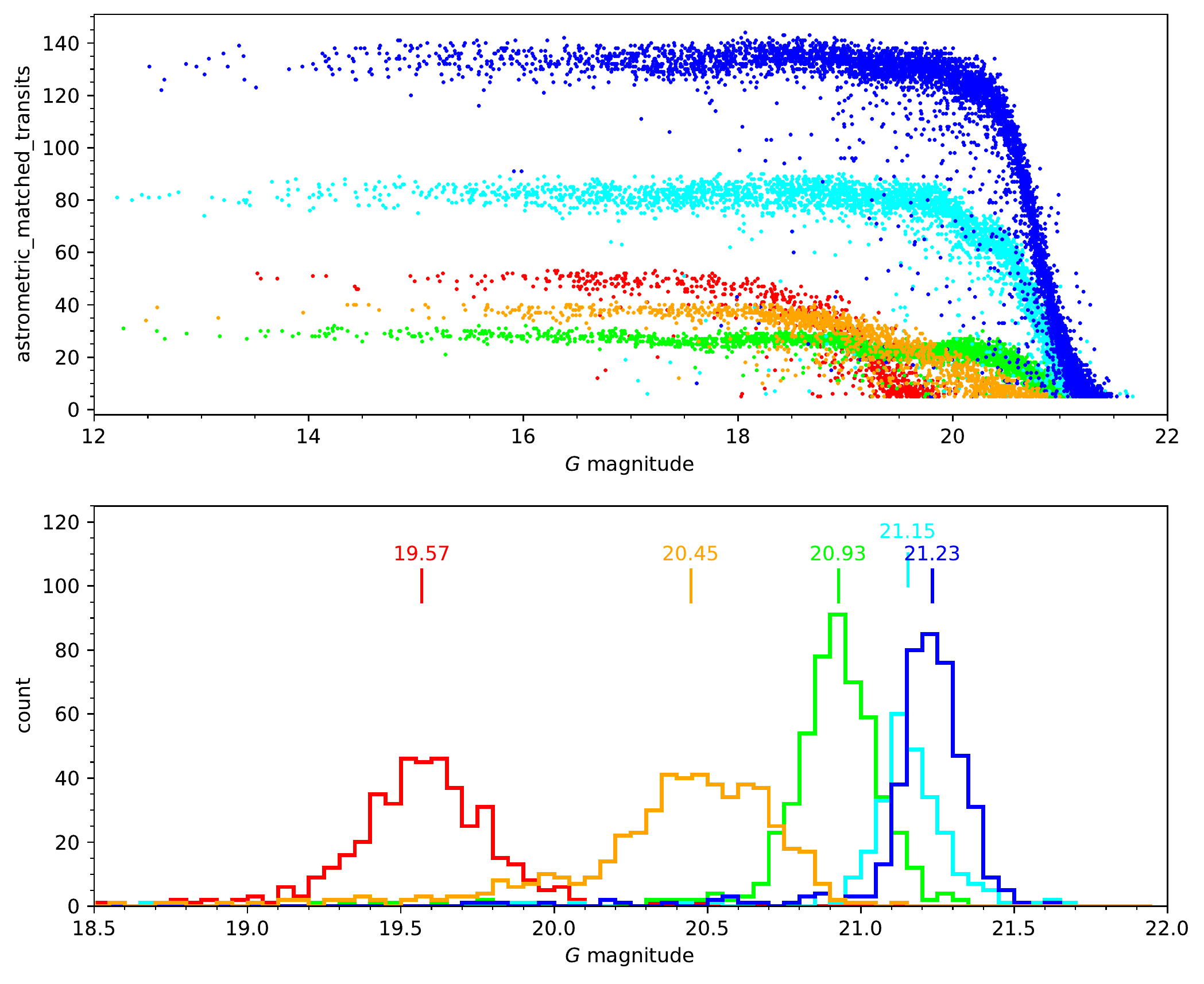}
    \caption{ Top: number of \texttt{astrometric\_matched\_transits} against magnitude for sources in five selected regions. Bottom: magnitude distribution of the sources with \texttt{astrometric\_matched\_transits} $\leq 10$, and their median magnitude $M_{10}$.
    }
    \label{fig:amt_vs_Gmag}
\end{figure*}

For the sources brighter than $G\sim18$, the number of transits matched to a given source is directly related to the number of times a region of the sky was observed by \Gaia. The number of matched transits decreases for fainter sources, which are not observed at every transits due to the on-board processing prioritising bright sources (top panel of Fig.~\ref{fig:amt_vs_Gmag}) until it reaches a minimum value of five. Sources with \texttt{astrometric\_matched\_transits} $<5$ are not included in the \Gaia catalogue. The $G$ magnitude at which this minimum value is reached varies across the sky, as it strongly depends on the level of crowding and on how many times the region was seen by \Gaia. 

In this paper, we note $M_{10}$ the median magnitude of the sources with \texttt{astrometric\_matched\_transits} $\leq 10$, which quantifies the typical magnitude of the sources. This quantity can be computed for any patch of sky and is a good predictor of completeness at all magnitudes.

\section{Location of our reference DECaPS fields}
\begin{figure*}
    \centering
    \includegraphics[width=0.99\textwidth]{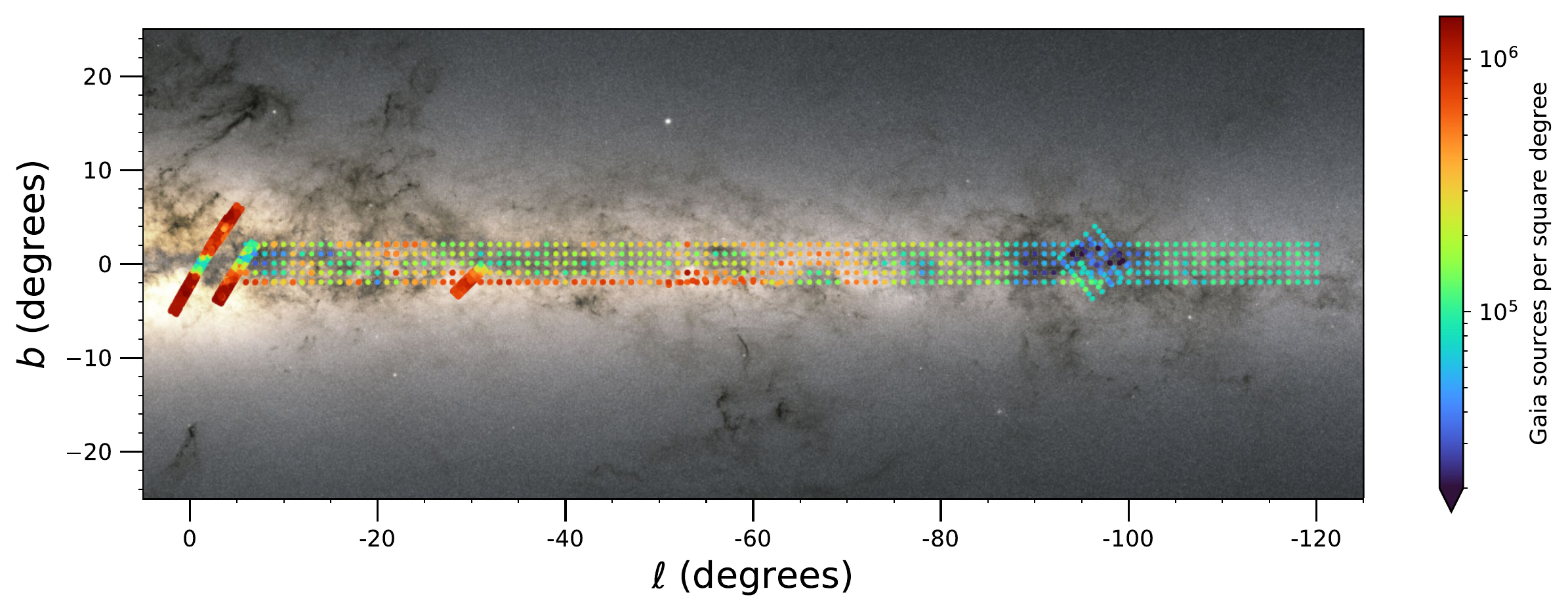}
    \caption{ Location of the 1085 patches of sky (some of which are further divided into smaller patches for a total of 2906) used to compute \Gaia-to-DECaPS source ratios and calibrate our model. Their observed (\Gaia) source densities range from $\sim$4000 to $\sim$1.4 million sources per square degree. The background map is the integrated \Gaia flux map (ESA/\Gaia DPAC).
    }
    \label{fig:map_decaps_fields}
\end{figure*}

\section{Sources without a $G$ magnitude}
\begin{figure*}
    \centering
    \includegraphics[width=0.99\textwidth]{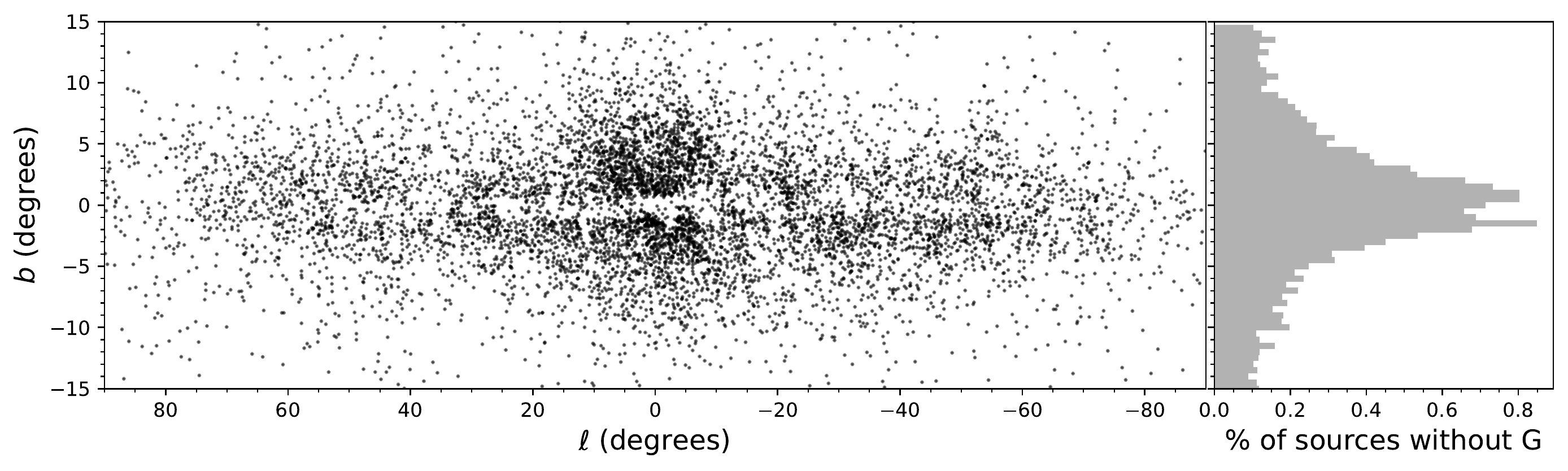}
    \caption{ Left: sky distribution of 100,000 randomly-selected \Gaia sources without a $G$ magnitude. The area displayed focuses on the inner Milky Way, but such sources are found all across the sky. Right: fraction of sources (out of a random sample of 3 million \Gaia sources) without a $G$ magnitude, computed in bins of Galactic latitude. The fraction is close to 0.3\% over the entire sky but is significantly higher in regions close to the Galactic plane.
    }
    \label{fig:map_noG}
\end{figure*}

This paper presents a selection function for the sample of \Gaia DR3 sources with published positions and magnitude $G$. A small fraction (variable across the sky but on average 0.3\% of the \Gaia DR3 catalogue) do not have associated $G$ magnitude. We illustrate their distribution in Fig.~\ref{fig:map_noG}.

\section{Generalised Sigmoid}

\begin{figure*}
    \centering
    \includegraphics[width=0.95\textwidth]{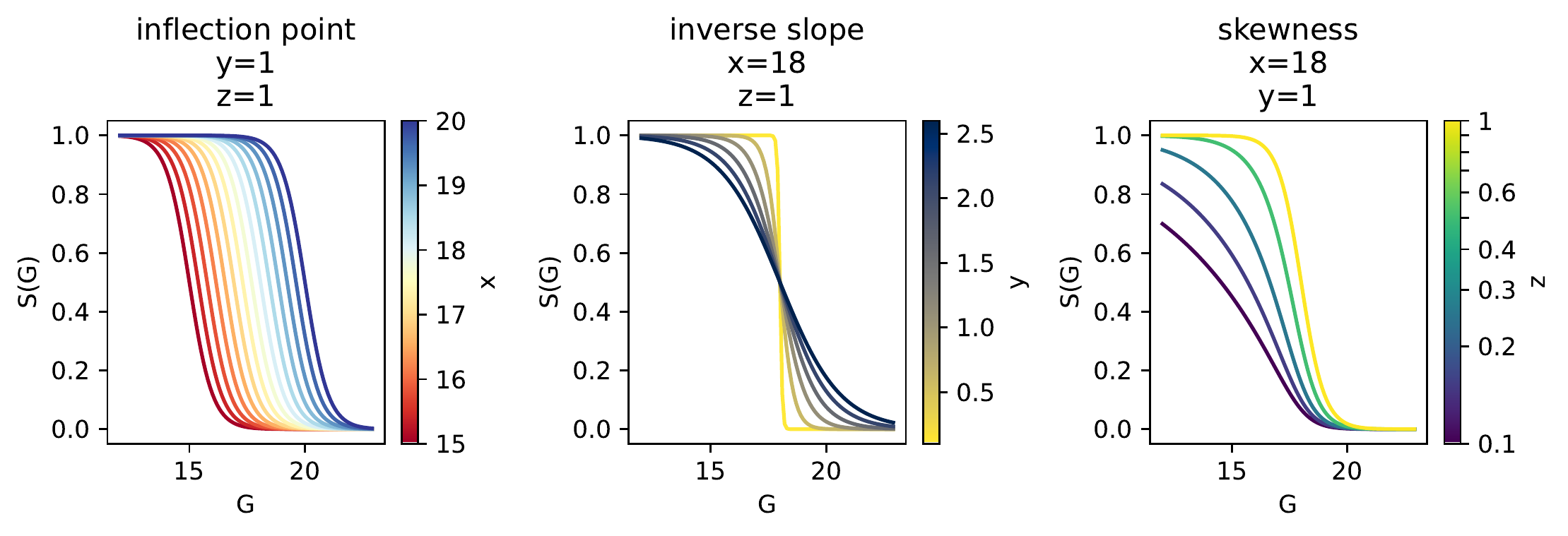}
    \caption{ Plots of the generalised sigmoid function defined in equation~\ref{eq:sigmoid} obtained varying the $x$, $y$, and $z$ parameters.
    }
    \label{fig:sigmoid_model}
\end{figure*}

\section{Model fitting}

The data we fit in this paper are the source count ratios of \Gaia to DECaPS, computed in 40 magnitude bins ($G$=15 to 23 in step of 0.2\,mag) in 2906 patches on the sky, representing a total of $n=40\times2906 = 116240$ data points $obs_i$.
The model presented in Sect.~\ref{sec:fitting} contains ten parameters $\Theta = (a_x,b_x,c_x,a_y,b_y,c_y,a_z,b_z,c_z,M_{break})$ and predicts the completeness for a given value of $M_{10}$ at a given magnitude $G$.

Assuming that the data (observed count ratios $obs_i$) are affected by Gaussian noise, the likelihood of the model (predicting completeness $pred_i$) is:

\begin{equation}
    \mathcal{L}(\Theta) = \prod_{i=1}^{n} \frac{1}{\sqrt{2 \pi \sigma ^2}} e^{-\frac{(obs_i-pred_i)^2}{2 \sigma^2}}
\end{equation}

\begin{figure*}
    \centering
    \includegraphics[width=0.95\textwidth]{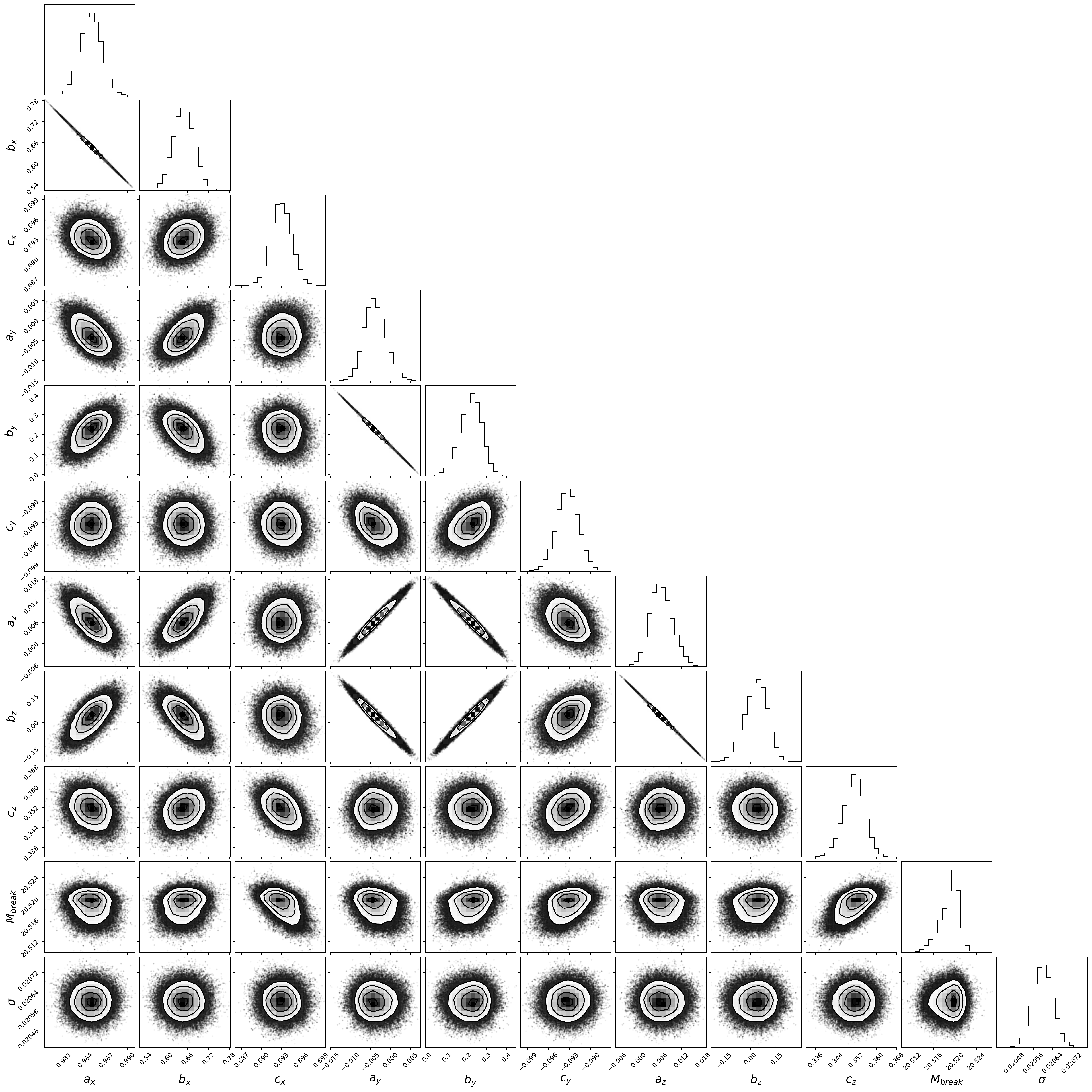}
    \caption{ Corner plot of the posterior chains of the ten parameters $(a_x,b_x,c_x,a_y,b_y,c_y,a_z,b_z,c_z,M_{break})$ constraining the selection function $S(G|M_{10})$, and the nuisance parameter $\sigma$.
    }
    \label{fig:cornerplot_11}
\end{figure*}

The corresponding log-likelihood is:
\begin{equation}
\begin{split}
    \log \mathcal{L}(\Theta) =& \sum_{i=1}^{n} \log \frac{1}{\sqrt{2 \pi \sigma ^2}} e^{-\frac{(obs_i-pred_i)^2}{2 \sigma^2}} \\
    =& n \log \frac{1}{\sqrt{2 \pi \sigma ^2}} - \frac{1}{2 \sigma^2} \sum_{i=1}^{n} (obs_i-pred_i)^2 \\
    =& n \log \sqrt{2 \pi} - n \log \sigma - \frac{1}{2 \sigma^2} \sum_{i=1}^{n} (obs_i-pred_i)^2
\end{split}
\end{equation}

\noindent which for a fixed value of $\sigma$ simplifies to ordinary least squares. Here we keep $\sigma$ as a free parameter, and maximise the quantity $\log \mathcal{L}(\Theta,\sigma)$. We adopt flat priors on all parameters, only enforcing that $\sigma$ must be positive. The corner plot of the posterior chain is shown in Fig.~\ref{fig:cornerplot_11}. We choose the median of each parameter chain as the best value for each parameter. These values are reported in table~\ref{tab:mcmc}.

In practice, the noise on the source count ratios varies with $M_{10}$ and $G$, is unlikely to be Gaussian, and is truncated since the observed completeness is bounded between 0 and 1 by design. The parameter $\sigma$ must therefore be considered as a nuisance parameter rather than a full model of the noise.

\newpage

\onecolumn
\section{Example use of our Python selection function package} \label{ap:python}

Here we show two examples of maps created with the model presented in this study, obtained through our Python package \texttt{gaiasf}. 
The package is hosted on Github\footnote{\url{https://github.com/gaia-unlimited/gaiasf}}. Further examples, including the construction of higher-resolution maps, are available as a Jupyter notebook tutorial\footnote{\url{https://github.com/gaia-unlimited/gaiasf/blob/main/docs/dr3-empirical-completeness.ipynb}}.

This section is not intended as documentation for the \texttt{gaiasf} package. Complete documentation and tutorials can be found through the Github repository and will be kept updated with every release of the code.

\subsection{All-sky map} \label{ap:ex0}

\begin{figure*}
    \centering
    \includegraphics[width=0.95\textwidth]{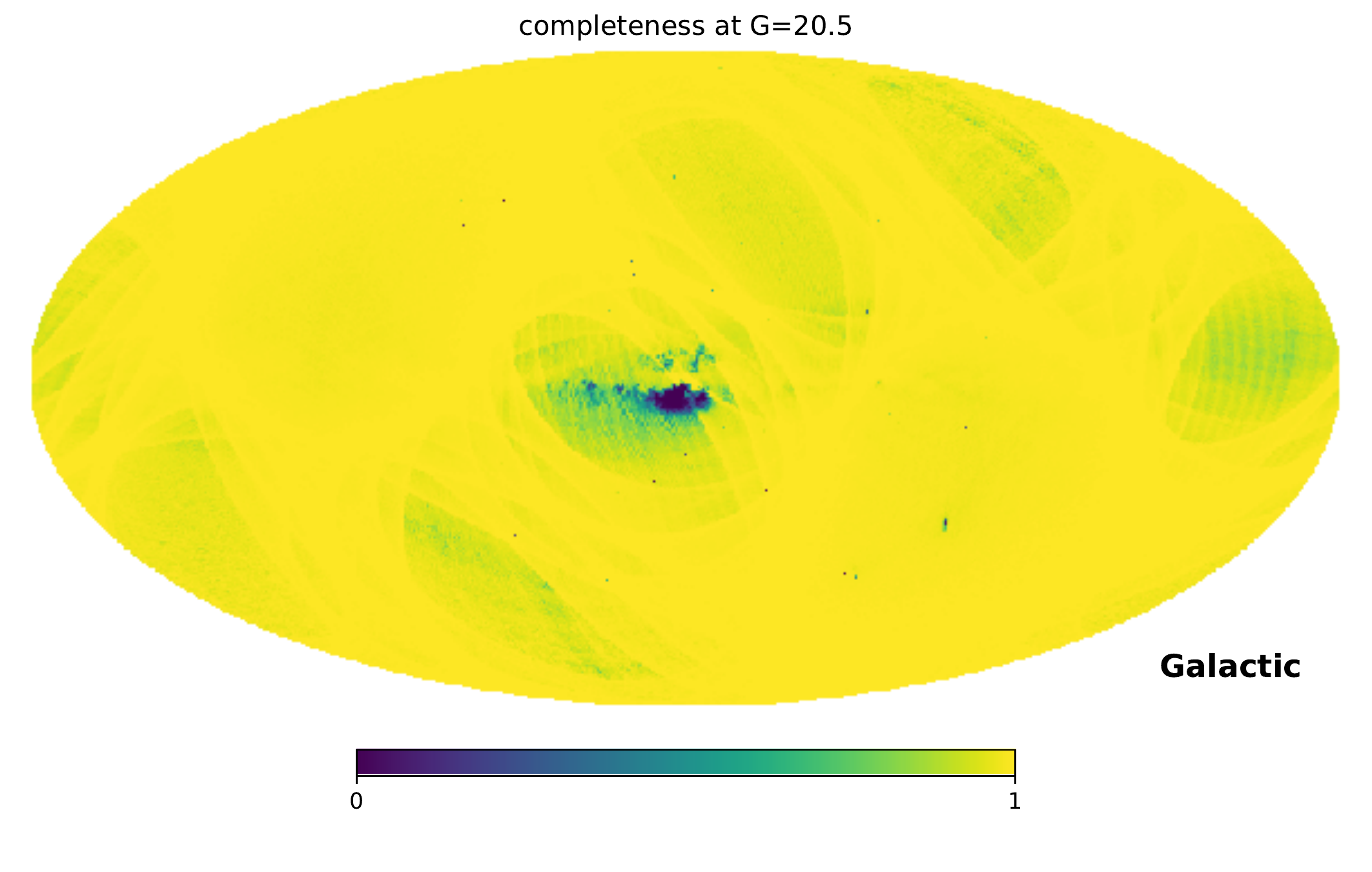}
    \caption{ Figure created by the coded provided in Sect.~\ref{ap:ex0}, using our \texttt{gaisf} package.
    }
    \label{fig:example0}
\end{figure*}

The following snippet of code produces the all-sky completeness map shown in Fig.~\ref{fig:example0}.

\begin{lstlisting}[language=Python,numbers=none]
from gaiasf.selectionfunctions import DR3SelectionFunctionTCG_hpx7
dr3sf = DR3SelectionFunctionTCG_hpx7()

from gaiasf.utils import get_healpix_centers
coords_of_centers = get_healpix_centers(7)
import numpy as np
gmag = np.ones_like(coords_of_centers) * 21
print(f'Computing the completeness for {len(coords_of_centers)} points.')
completeness = dr3sf.query(coords_of_centers,gmag)

import healpy as hp
hp.mollview(completeness, coord=['Celestial','Galactic'], min=0,max=1, title='completeness at G=21')

\end{lstlisting}

\newpage

\subsection{Grid of coordinates} \label{ap:ex1}

\begin{figure*}
    \centering
    \includegraphics[width=0.95\textwidth]{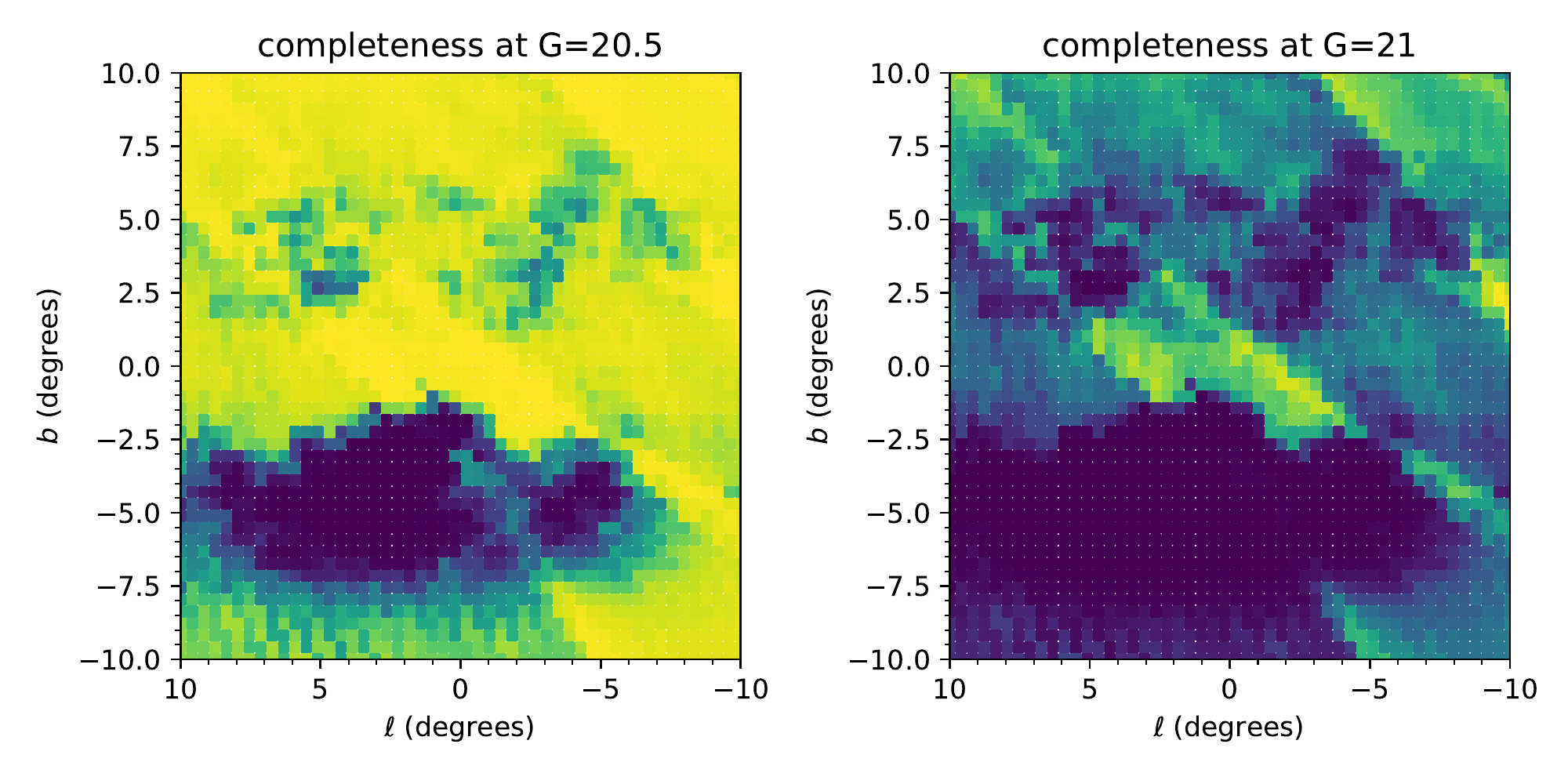}
    \caption{  Figure created by the provided shown in Sect.~\ref{ap:ex1}, using our \texttt{gaisf} package.
    }
    \label{fig:example1}
\end{figure*}

The following snippet of code produces Fig.~\ref{fig:example1}.

\begin{lstlisting}[language=Python,numbers=none]
from gaiasf.selectionfunctions import DR3SelectionFunctionTCG_hpx7
dr3sf = DR3SelectionFunctionTCG_hpx7()

import numpy as np
from astropy import units as u
from astropy.coordinates import SkyCoord

l = np.linspace( -10 , 10 , 50)
b = np.linspace( -10, 10 , 50)
l, b = [c.flatten() for c in np.meshgrid(l, b)]
coords = SkyCoord(frame="galactic",l=l*u.degree, b=b*u.degree)

import matplotlib.pyplot as plt
plt.figure(figsize=(8,4))
for i,G in enumerate([20.5,21]):
    completeness = dr3sf.query(coords,G*np.ones_like(coords))
    plt.subplot(1,2,i+1)
    plt.scatter( l , b , c=completeness,s=12,marker='s',vmin=0,vmax=1)
    plt.xlabel('$\ell$ (degrees)'); plt.ylabel('$b$ (degrees)')
    plt.title(f'completeness at G={G}')
    plt.xlim(10,-10); plt.ylim(-10,10)
    plt.minorticks_on()
plt.tight_layout()


\end{lstlisting}

\end{document}